\documentclass[12pt,letterpaper]{article}

\usepackage{hyperref}

\usepackage{amsmath}
\allowdisplaybreaks[1]

\usepackage{graphicx}
\usepackage{subfig}
\graphicspath{{fig/}}

\usepackage{amssymb}

\usepackage{afterpage}

\usepackage{multirow}
\usepackage{rotating}
\usepackage{array}

\usepackage{listings}
\usepackage{color}
\definecolor{dkgreen}{rgb}{0,0.4,0}
\definecolor{gray}{rgb}{0.5,0.5,0.5}

\usepackage{longtable}

\def\be{\begin{equation}}
\def\ee{\end{equation}}
\def\ba{\begin{align}}
\def\ea{\end{align}}

\newcommand{\eq}[1]{(\ref{#1})}

\def\eqdef{\overset{\text{def}}{=}}

\newcommand{\comment}[1]{}

\newcommand\E[1]{\cdot 10^{#1}}
\newcommand{\code}[1]{{#1}} 

\newcommand{\fln}{\left\lfloor n/2\right\rfloor}

\begin{document}


\begin{titlepage}
\vskip 1in
\begin{center}
{\Large
{Time-Symmetric Rolling Tachyon Profile}}
\vskip 0.5in
{Matheson Longton}
\vskip 0.3in
{\it 
Department of Physics and Astronomy\\
University of British Columbia\\
Vancouver, Canada}
\end{center}

\vskip 0.5in
\begin{abstract}
We investigate the tachyon profile of a time-symmetric 
rolling tachyon solution to open string field theory. We 
algebraically construct the solution of \cite{Kiermaier:2007vu} at 
6th order in the marginal parameter, and numerically evaluate the 
corresponding tachyon profile as well as the action and several 
correlation functions containing the equation of motion.  We find 
that the marginal operator's singular self-OPE is properly 
regularized and all quantities we examine are finite. In contrast to 
the widely studied time-asymmetric case, the solution depends 
nontrivially on the strength of the deformation parameter.  For 
example, we find that the number and period of oscillations of the 
tachyon field changes as the strength of the marginal deformation is 
increased. We use the recent renormalization scheme of 
\cite{Karczmarek:2014wea}, which contains two free parameters. At 
finite deformation parameter the tachyon profile depends on these 
parameters, while when the deformation parameter is small, the 
solution becomes insensitive to them and behaves like previously 
studied time-asymmetric rolling tachyon solutions. We also show that 
convergence of perturbation series is not as straightforward as in 
the time-asymmetric case with regular OPE, and find evidence that it 
may depend on the renormalization constants. 
\end{abstract}
\end{titlepage}





\section{Introduction and Conclusions}

In a boundary CFT, the boundary condition can be deformed on any 
section of the boundary by exponentiating a marginal operator 
integrated along it, as in 
\be
e^{\lambda\int dt\,V(t)}~.
\ee
The marginal parameter $\lambda$ controls the strength of the 
deformation.  In Open String Field Theory, allowed D-brane 
configurations are in one to one correspondence with classical 
solutions.  The rolling tachyon is the time-dependent 
solution which corresponds to a decaying D-brane.  There are two 
rolling tachyon solutions obtained by different marginal deformations 
of the D-brane CFT.  The simpler case, the exponential rolling 
tachyon, involves the marginal operator 
$V(t)=e^{X^{0}(t)}$ and represents a D-brane which exists in the 
infinite past and then decays at a finite time.  This case has been 
studied using level truncation methods 
\cite{Coletti:2005zj,Erler:2004sy} as well as analytically 
\cite{Kiermaier:2007ba,Schnabl:2007az,Kiermaier:2010cf}, and is 
relatively simple because the OPE of the marginal operator with 
itself is finite.  The more difficult case uses the time-symmetric 
marginal operator $V=\sqrt{2}\cosh(X^{0})$, which has the singular 
self-OPE $V(0)V(x)\sim\frac{1}{x^{2}}$.  This rolling tachyon 
corresponds to placing a D-brane at $t=0$ and letting it decay at 
both $t=-\infty$ and $t=+\infty$.  

In SFT, the tachyon profile is the tachyon component of the string 
field as a function of time.  In the symmetric case it has the form 
\be \label{eq.rtc.Tx}
T(t)=2\sum_{n=1}^{\infty}\sum_{j=0}^{\fln}\lambda^{n}\beta_{n}^{(j)}\cosh((n-2j)t)~,
\ee
where $\beta_{n}^{(j)}$ are coefficients which can be 
calculated numerically and $\fln$ is the greatest integer less than 
or equal to $n/2$, so that $n-2j\geq0$.  
In this notation the deformation strength 
$\lambda$ is taken to be negative for physical solutions 
\cite{Kiermaier:2007ba}.  In the exponential case, $\lambda$ controls 
the time at which the D-brane decays, while for the time-symmetric 
case it determines the lifetime of the 
D-brane, with longer lifetimes corresponding to $\lambda$ closer to 
zero.  In the regular OPE case, instead of the double sum and 
time-symmetric $\cosh$ functions, energy conservation 
tells us that there is only a single sum of exponentials involving 
coefficients with $\beta_{n}^{(0)}$:
\be
T_{\text{reg}}(t)=\sum_{n=1}^{\infty}\lambda^{n}\beta_{n}^{(0)}e^{nt}~.
\ee
While the $\beta_{n}^{(j)}$ are in general 
gauge-dependent, for one choice of gauge it was proven that the 
sum in the regular OPE case converges for all $\lambda$, with the 
asymptotic behaviour $\beta_{n}^{(0)}\sim e^{-\gamma n^{2}}$ 
\cite{Kiermaier:2010cf}. Numerical data suggests that this is true in 
other gauges as well, as shown in figure 
\ref{fig.rtc.edge-gamma-all}.  The trouble with this is that 
the tachyon profile itself exhibits wild oscillations which grow 
exponentially in magnitude, while the vacuum without any D-branes is 
a well defined and finite point in string field space.  How these two 
very different looking string fields are reconciled has been the 
source of much speculation (see, for example, 
\cite{Ellwood:2007xr}).  Our results confirm that the tachyon profile 
has the same growing oscillatory behaviour in the time-symmetric 
case, and do not appear to exclude any of the current hypotheses.

When studying the marginal operator $V=\sqrt{2}\cosh(X^{0})$ leading 
to the time-symmetric rolling tachyon, we must be careful to avoid 
singularities arising from the operator's OPE.  Analytic solutions 
for marginal deformations require the insertion of many copies of the 
marginal operator with separations that are integrated over, and 
there will be divergences when two operators approach each other.  
Fortunately there are several solutions which are intended to handle 
this issue \cite{Kiermaier:2007vu, Fuchs:2007yy, Maccaferri:2014cpa, Erler:2014eqa}.  The 
most recent work, by Erler and Maccaferri, does not apply to 
solutions which have a non-trivial time direction, so we cannot use 
it for the rolling tachyon.  Fuchs, Kroyter and Potting's solution 
was designed with the photon marginal deformation in mind, but it is 
possible that it could describe the rolling tachyon as well.  The 
solution of \cite{Maccaferri:2014cpa} is a generalization of 
\cite{Kiermaier:2010cf} to operators with singular OPE, and it could 
be applied to the rolling tachyon.  In fact it has been suggested 
that this solution could give the tachyon profile in the form 
\eq{eq.rtc.Tx-betaeff}, which would help settle the convergence issue.

Our focus, however, will be on the work of Kiermaier and Okawa.  They 
proposed a general construction dependent on the existence of a 
suitable renormalization scheme \cite{Kiermaier:2007vu}, which was 
investigated and refined 
in \cite{Karczmarek:2014wea}.  A general renormalization scheme 
satisfying the necessary conditions was shown in 
\cite{Karczmarek:2014wea} to have at least two 
free parameters, suggesting that the tachyon profile could have free 
parameters as well.  Here we will perform the first explicit 
numerical calculations for this solution, and we will show that the 
tachyon profile is a finite function which does in fact depend on the 
free parameters.

When we implement the solution $\Psi$ of \cite{Kiermaier:2007vu} with the 
renormalization scheme of \cite{Karczmarek:2014wea} we can find the 
tachyon profile for the symmetric rolling tachyon.  This involves 
algebraically constructing the wedge states with insertions 
corresponding to the solution, taking expectation values, and then 
performing the required integrals numerically.  Here this is done 
up to 6th order in $\lambda$.  It will have the form \eq{eq.rtc.Tx}, 
where now the function is symmetric in $t$ and all the 
$\beta_{n}^{(j)}$ are non-zero.  The marginal operator 
$\sqrt{2}\cosh(X^{0})$ contains the operators $e^{\pm X^{0}}$ with 
both signs, and the coefficients $\beta_{n}^{(j)}$ correspond to 
terms with $n-j$ factors of one of the two operators and $j$ factors 
of the other.  Since renormalization 
has the effect of adding counterterms for collisions of operators 
with opposite sign, the 
$j=0$ coefficients involve no counterterms and behave very 
similarly to exponential solutions.  These show the same 
$\beta_{n}^{(0)}\sim e^{-\gamma n^{2}}$ asymptotic 
behaviour, implying that the sum 
$\sum_{n=1}^{\infty}\lambda^{n}\beta_{n}^{(0)}\cosh(nt)$ converges 
absolutely for all $\lambda$.
For $|\lambda|\ll1$, as is the case when the D-brane survives for a 
long time, the $j>0$ coefficients are suppressed due to extra factors 
of $\lambda$.  This 
results in a decay process which looks very much like the regular 
case, as the decay is well separated from the ``anti-decay'' by the 
D-brane's lifetime.  Once this lifetime is long enough, further 
shrinking $\lambda$ even has the same effect on the decay time as it 
would with the exponential rolling tachyon, simply shifting the time 
of the decay.

For the $\beta_{n}^{(j)}$ coefficients with $j>0$, each coefficient is 
calculated using a number of counterterms determined by $j$.  The 
counterterms in turn are functions of the two parameters of the 
renormalization scheme, $C_{0}$ and $C_{1}$.  The bulk coefficients 
are therefore polynomial functions of $C_{0}$ and $C_{1}$.  
Because these coefficients are not constants, patterns such as the 
asymptotic behaviour for $j=0$ could depend on the choice of $C_{0}$ 
and $C_{1}$.  Considering only the $j=1$ coefficients, with 
$C_{0}=C_{1}=0$ they are quite a good fit to 
$\beta_{n}^{(1)}\sim e^{-\gamma_{1}(n-2)^{3}}$.  In fact there is no 
choice of those constants for which the exponential quadratic 
behaviour $\beta_{n}^{(1)}\sim e^{-\gamma_{1}(n-2)^{2}}$ fits as 
closely.  This suggests that the sum 
$\sum_{n=2}^{\infty}\lambda^{n}\beta_{n}^{(1)}\cosh((n-2)t)$ also 
converges, but there may still be some choices of $C_{0}$ and $C_{1}$ 
for which this is not the case, or for which the radius of 
convergence in $\lambda$ is finite.  For example, choosing the 
constants so that $\beta_{n}^{(1)}$ are a best fit to the 
exponential cubic behaviour results in $\beta_{n}^{(2)}$ which are 
increasing, at least for the three coefficients we can calculate with 
$j=2$.  

So how does the inclusion of all the $\beta_{n}^{(j)}$ coefficients 
affect the shape of the tachyon profile?  We show that for 
$|\lambda|\ll1$ these coefficients are negligible, 
but as the D-brane lifetime is decreased there comes a point where 
more coefficients must be considered.  Some terms cease to dominate for 
any range of time, and the number of oscillations actually 
decreases.  The missing oscillation means that the effective 
``period'' is significantly decreased.  What this means physically is 
not clear, since the period is a gauge dependent quantity related to 
the coefficient $\gamma$ in the exponent of the asymptotic 
behaviour.  The tachyon profile for small $\lambda$ is very similar to 
that of \cite{Kiermaier:2007ba}, while for large $\lambda$ it has 
features similar to the tachyon profile of \cite{Kiermaier:2010cf}, 
so perhaps the solution is interpolating between 
regular-OPE solutions in different gauges as the marginal deformation 
strength is changed.  Understanding this phenomenon is left for 
future work.

This paper is organized as follows.  Section \ref{sec.rtc.Tx} is 
focused on the details of the tachyon profile, and contains discussion 
and plots of results.  In section 
\ref{sec.rtc2.details} we briefly discuss how the numerical calculations 
were performed, and demonstrate that our results converge to 
appropriate values.  There we also discuss sources of roundoff error 
which can influence the calculations.

\section{The Tachyon Profile} \label{sec.rtc.Tx}

The solution of \cite{Kiermaier:2007vu} presents a promising 
framework for construction of a time-symmetric rolling tachyon 
solution, but it was not applied to any specific marginal 
deformation.  Taking that approach and inserting the marginal 
deformation $V=\sqrt{2}\cosh(X^{0})$ we are able to numerically 
compute the tachyon profile up to 6th order in the deformation 
parameter $\lambda$.  Since the tachyon profile has previously been 
calculated for several exponential rolling tachyon solutions, we can 
compare our results in order to determine what qualitative 
differences appear in the time-symmetric case.  It is also useful to 
have explicit numerical evidence that the renormalization scheme 
studied in \cite{Karczmarek:2014wea} is effective and the solution 
remains finite despite the singular OPE that the marginal operator 
has with itself.

The solution takes the form of a wedge state with insertions on the 
boundary.  While one insertion will always be at a fixed 
location, the rest are integrated.  The renormalization procedure 
replaces pairs of operators with appropriate counterterms under the 
integral.  Each operator $V$ contains two terms carrying $\pm1$ 
unit of ``momentum'' in the time direction, but the counterterms are 
functions and carry no momentum.  In \eq{eq.rtc.Tx} the coefficient 
$\beta_{n}^{(j)}$ clearly contains the part of the tachyon profile 
with $n$ factors of $\lambda$ and $k=n-2j$ units of this momentum, so 
it follows that the 
coefficients $\beta_{n}^{(0)}$ contain no counterterms.  This is as 
it should be since operators $e^{\pm X^{0}}$ with the same sign have 
a regular OPE; the singular OPE of the $\cosh(X^{0})$ marginal 
operator comes entirely from the collision of exponentials with 
opposite sign.  The index $j$, which counts the momentum deficit, 
also has the effect of counting the maximum number of counterterm 
factors.  For the coefficients $\beta_{n}^{(j)}$, table 
\ref{tab.rtc.solutionmodesRC} shows their values as calculated by the 
\code{Cuhre} algorithm, and with the exception of two terms we will 
use those coefficients.  For technical reasons explained in section 
\ref{sec.rtc.checks}, the two terms marked with asterisks will use 
values found by the \code{Suave} algorithm instead, and those are 
shown in table \ref{tab.rtc.solutionmodesS}.
Occasionally we will want to think of the 
tachyon profile 
in terms of these timelike momentum modes, and write 
\be \label{eq.rtc.Tx-ksum}
T(t)=\sum_{k=0}^{\infty}2\cosh(kt)\sum_{n=k}^{\infty}\lambda^{n}\beta_{n}^{\left(\frac{n-k}{2}\right)}~.
\ee
This form is equivalent to \eq{eq.rtc.Tx} as long as we define 
$\beta_{n}^{(j)}$ to vanish for non-integer $j$, as well as for 
$n=j=0$. 


\begin{sidewaystable}[htbp]\centering\begin{tabular}{|cc|>{$}l<{$}|}
\hline
$n$&$j$&\multicolumn{1}{c|}{$\beta_{n}^{(j)}$} \\
\hline
1&\bf 0&\frac{1}{\sqrt{2}} \\
2&1&(-1.29904\ldots \pm 3\E{-11})+(0\pm1\E{-14})C^{L} \\
2&\bf 0&(0.0760297\ldots \pm 8\E{-16}) \\
3&1&(-1.30572\pm4.3\E{-5})-(0.707107\ldots\pm3\E{-14})C_{1}-(0\pm2\E{-4})C^{L} \\
3&\bf 0&(9.150\pm0.019)\E{-4} \\
4&2&(0.655579\pm6\E{-6})+(0\pm3\E{-4})C^{L}+(3.2858\pm0.0021)C_{1}+(4.9\pm7.8)\E{-15}C^{L}C_{1} \\
	&&\quad+(0\pm1\E{-3})C_{0}+(0\pm1\E{-14})C^{L}C_{0} \\
4&1&-(0.4488\pm0.0031)+(0\pm8\E{-4})C^{L}-(0.2349\pm0.0023)C_{1}+(1.4\pm0.7)\E{-7}C_{0} \\
4&\bf 0&(1.17222\pm0.00013)\E{-6} \\
5&2&(0.723\pm0.011)+(0\pm1\E{-3})C^{L}+(4.387\pm0.041)C_{1}+(0\pm0.02)C^{L}C_{1}+(3.53553\ldots\pm7\E{-15})C_{1}^{2} \\
	&&\quad+(0\pm6\E{-3})C_{0}+(0\pm0.01)C^{L}C_{0} \\
5&1&(-0.01221\pm1.2\E{-4})+(0\pm2\E{-5})C^{L}-(5.86\pm0.34)\E{-3}C_{1}-(1.27\pm0.61)\E{-4}C_{0} \\
5&\bf 0&(1.598\pm0.007)\E{-10} \\
6&3&(-0.3\pm0.4)^{*}+(0\pm3\E{-3})C^{L}-(2.572\pm0.026)C_{1}+(0.3\pm1.2)\E{-3}C^{L}C_{1}-(23.9401\pm0.0013)C_{1}^{2} \\
	&&\quad+(1.5\pm3.1)\E{-14}C^{L}C_{1}^{2}+(0.0955\pm0.0030)C_{0}+(0\pm5\E{-3})C^{L}C_{0}-(0.135\pm0.015)C_{0}C_{1} \\
	&&\quad-(1.2\pm3.9)\E{-14}C^{L}C_{0}C_{1}+(5.8\pm1.5)\E{-6}C_{0}^{2}\\
6&2&(0.4991\pm0.0050)+(0.4\pm1.3)\E{-5}C^{L}+(1.912\pm0.019)C_{1}+(0.8\pm4.1)\E{-3}C^{L}C_{1}+(1.715\pm0.024)C_{1}^{2} \\
	&&\quad+(1.879\pm0.025)\E{-2}C_{0}+(0\pm2\E{-3})C^{L}C_{0}+(4.77\pm0.38)\E{-2}C_{0}C_{1}-(2.37\pm0.09)\E{-7}C_{0}^{2}\\
6&1&(-2.686\pm0.027)\E{-5}-(1.6\pm5.9)\E{-8}C^{L}-(9.1\pm2.2)\E{-6}C_{1}-(7.3\pm0.7)\E{-7}C_{0} \\
6&\bf 0&(2.18\pm0.04)\E{-15\,*} \\
\hline
\end{tabular}\caption[Deterministic numerical results for the 
time-symmetric rolling tachyon profile]
{The non-zero coefficients $\beta_{n}^{(j)}$ of the tachyon profile 
for the $\cosh$ rolling tachyon with singular self-OPE.  
\code{Cuhre}/\code{QAG} results shown.  The constant $C^{L}$ is part 
of the renormalization scheme of \cite{Karczmarek:2014wea}, but it 
can not influence the solution, so we safely set it to zero in our 
analysis.  $C^{L}$ was included in these numerical results only to 
demonstrate that it does not contribute to the solution at all.\\
${}^{*}$ These two coefficients 
found using the Cuhre algorithm appear to be unreliable, so the 
corresponding Suave results in table \ref{tab.rtc.solutionmodesS} 
will be used for analysis instead.}
\label{tab.rtc.solutionmodesRC}\end{sidewaystable}

\begin{sidewaystable}[htbp]\centering\begin{tabular}{|cc|>{$}l<{$}|}
\hline
$n$&$j$&\multicolumn{1}{c|}{$\beta_{n}^{(j)}$} \\
\hline
1&\bf 0&\frac{1}{\sqrt{2}} \\
2&1 & -(1.2985\pm0.0003) - (4.134\pm0.007)\E{-6}C^L \\
2&\bf 0 & (7.61\pm0.07)\E{-2} \\
3&1 & -(1.301\pm0.005) - (0.001\pm0.010)C^L - (0.707107\ldots\pm7\E{-18})C_1 \\
3&\bf 0 & (8.99\pm0.09)\E{-4} \\
4&2 & (0.659\pm0.007) + (0.2\pm2.6)\E{-3}C^L + (3.288\pm0.003)C_1 + (0\pm1\E{-17})C_1C^L + (0.5\pm3.4)\E{-3}C_0 \\
	&&\quad + (0\pm5\E{-9})C_0C^L \\
4&1 & -(0.449\pm0.004) - (0.1\pm1.4)\E{-3}C^L - (0.235\pm0.002)C_1 + (1.39\pm0.07)\E{-4}C_0 \\
4&\bf 0 & (1.163\pm0.002)\E{-6} \\
5&2 & (0.722\pm0.012) + (1.3\pm1.3)\E{-3}C^L + (4.38\pm0.04)C_1 + (0.014\pm0.034)C_1C^L + (3.53553\ldots\pm3\E{-8})C_1^2 \\
	&&\quad + (0.1\pm6.5)\E{-3}C_0 - (0.3\pm1.3)\E{-2}C_0C^L \\
5&1 & -(1.21\pm0.01)\E{-2} + (2.2\pm1.1)\E{-5}C^L - (5.81\pm0.06)\E{-3}C_1 - (1.17\pm0.01)\E{-4}C_0 \\
5&\bf 0 & (1.27\pm0.01)\E{-10} \\
6&3 & -(0.307\pm0.004) - (1.6\pm2.9)\E{-3}C^L - (2.55\pm0.05)C_1 + (0.5\pm3.6)\E{-2}C_1C^L - (23.943\pm0.008)C_1^2 \\
	&&\quad + (0\pm1\E{-17})C_1^2C^L + (9.5\pm0.5)\E{-2}C_0 - (8.0\pm6.6)\E{-3}C_0C^L - (0.12\pm0.02)C_0C_1 \\
	&&\quad + (0\pm1\E{-17})C_0C_1C^L + (5.0\pm4.4)\E{-5}C_0^2 \\
6&2 & (0.497\pm0.005) - (2.1\pm1.1)\E{-4}C^L + (1.92\pm0.02)C_1 - (1.7\pm2.6)\E{-3}C_1C^L + (1.718\pm0.013)C_1^2 \\
	&&\quad+ (1.774\pm0.014)\E{-2}C_0 - (1.2\pm2.3)\E{-3}C_0C^L + (4.71\pm0.05)\E{-2}C_0C_1 - (1.156\pm0.016)\E{-4}C_0^2 \\
6&1 & -(2.54\pm0.02)\E{-5} - (3.5\pm0.3)\E{-8}C^L - (1.222\pm0.012)\E{-5}C_1 - (7.618\pm0.014)\E{-7}C_0 \\
6&\bf 0 & (2.3\pm0.3)\E{-15} \\
\hline
\end{tabular}\caption[Monte Carlo numerical results for the 
time-symmetric rolling tachyon profile]
{The non-zero coefficients $\beta_{n}^{(j)}$ of the tachyon profile 
for the $\cosh$ rolling tachyon with singular self-OPE.
\code{Suave} results shown for comparison with the deterministic 
results of table \ref{tab.rtc.solutionmodesRC}.}
\label{tab.rtc.solutionmodesS}\end{sidewaystable}

\comment{ 
The coefficients are shown in table \ref{tab.rtc.solutionmodesRC}.  
Since the tachyon profile is gauge dependent, we cannot say that in 
some limit it should approach the exponential case, so in order to 
judge the accuracy of the numerical integration we can compare the 
results using two different numerical 
integration algorithms.  We perform all calculations using both a 
deterministic algorithm, \code{Cuhre}, and an adaptive Monte Carlo 
algorithm, \code{Suave}.  
Since using too few samples causes wildly erratic results for each 
integral, the 93 \code{Cuhre} integrals used in calculating the 
coefficients of table \ref{tab.rtc.solutionmodesRC} were each 
compared to the corresponding \code{Suave} results.  
After identifying several integrals deserving of further scrutiny, 
higher precision calculations yielded deterministic results which 
were reliable in all but two cases.  These two integrals, for which 
we will use \code{Suave} results instead of \code{Cuhre}, contribute 
to $\beta_{6}^{(0)}$ and $\beta_{6}^{(3)}$ and are marked with 
asterisks in table \ref{tab.rtc.solutionmodesRC}.  A full set of 
\code{Suave} results for the tachyon profile is given in table 
\ref{tab.rtc.solutionmodesS}.
}

The tachyon profile for several different solutions with 
regular OPE has been calculated before.  It has the simpler form of 
$T(t)=\sum_{n=0}^{\infty}\lambda^{n}\beta_{n}\sqrt{2}^{n}e^{nt}$ 
where the coefficients $\beta_{n}\eqdef\beta_{n}^{(0)}$ are only 
non-zero for maximal momenta.  In table \ref{tab.rtc.korzmodes} we 
compare the coefficients for those solutions to the 
ones we have found.  We have changed the normalization of their 
coefficients by $2^{-\frac{n}{2}}$ for better comparison with our 
coefficients, due to the relative normalizations of the marginal 
operators $e^{X^{0}}$ and $\sqrt{2}\cosh(X^{0})$.  Our coefficients 
show very similar falloff to \cite{Kiermaier:2007ba} as $n$ is 
increased, though we do not expect exact agreement between 
any of the sets of coefficients because the tachyon profile is a gauge 
dependent quantity.  We believe that each of these lists is related 
to the others by such gauge transformations, but constructing them is 
beyond the scope of this work. 

\begin{table}[htb]\centering\begin{tabular}{|c|c|c|c|c|}
\hline & \cite{Kiermaier:2007ba} & \cite{Kiermaier:2010cf} & \cite{Coletti:2005zj} & $\Psi$ here \\
\hline $n$ & \multicolumn{4}{c|}{$\beta_{n}$} \\
\hline 1 & $\frac{1}{\sqrt{2}}$ & $\frac{1}{\sqrt{2}}$ & $\frac{1}{\sqrt{2}}$ & $\frac{1}{\sqrt{2}}$ \\
2 & 0.0760295 & 0.290 & 0.0760297 & 0.0760297 \\
3 & $7.59312\E{-4}$ & 0.0506 & $7.732\E{-4}$ & $9.149\E{-4}$ \\
4 & $6.54812\E{-7}$ & $4.18\E{-3}$ & $9.8145\E{-7}$ & $1.173\E{-6}$ \\
5 & $4.93424\E{-11}$ & $1.54\E{-4}$ & $8.734\E{-11}$ & $1.275\E{-10}$ \\
6 & $3.50136\E{-16}$ & $2.45\E{-6}$ & $7.903\E{-13}$ & $2.26\E{-15}$ \\
7 & $2.41180\E{-22}$ & $1.64\E{-8}$ && \\
\hline
\end{tabular}\caption[Comparison of the tachyon profile for several 
different solutions representing the exponential rolling tachyon]
{The coefficients for the tachyon modes of a 
purely exponential rolling tachyon with regular self-OPE.  Calculated 
from the solutions of \cite{Kiermaier:2007ba}, 
\cite{Kiermaier:2010cf}, and \cite{Coletti:2005zj}.  The $n=k$ 
coefficients for our calculations based on \cite{Kiermaier:2007vu} 
are included for comparison.}
\label{tab.rtc.korzmodes}\end{table}

As in \cite{Kiermaier:2007ba}, we take $\lambda$ to be negative in 
order to study physical solutions.  With this assumption, the tachyon 
profile \eq{eq.rtc.Tx} can be rewritten as 
\be \label{eq.rtc.Tx-ln}
T(t)=\sum_{n=1}^{\infty}\sum_{j=0}^{\fln}(-1)^{n}\beta_{n}^{(j)}\left(e^{n(\ln|\lambda|+t)-2jt}+e^{n(\ln|\lambda|-t)+2jt}\right)~,
\ee
where in practice the sum over $n$ only runs up to some cutoff $N$ 
where the coefficients can be computed.  When only the $j=0$ coefficients 
and the first term in parentheses are considered, as in the regular OPE 
case, we can clearly see that a change of $\ln|\lambda|$ will only 
shift the time of the D-brane decay.  For the singular case, however, 
the tachyon profile will have a different shape depending on the 
strength of the marginal deformation, controlled by $\ln|\lambda|$.  
The renormalization scheme also contains the constants $C_{0}$ and 
$C_{1}$, which will appear nontrivially in $\beta_{n}^{(j)}$ with 
$j>0$.  

\subsection{Small $\lambda$}
\begin{figure}
	\subfloat[]
	{\includegraphics[width=0.43\textwidth]{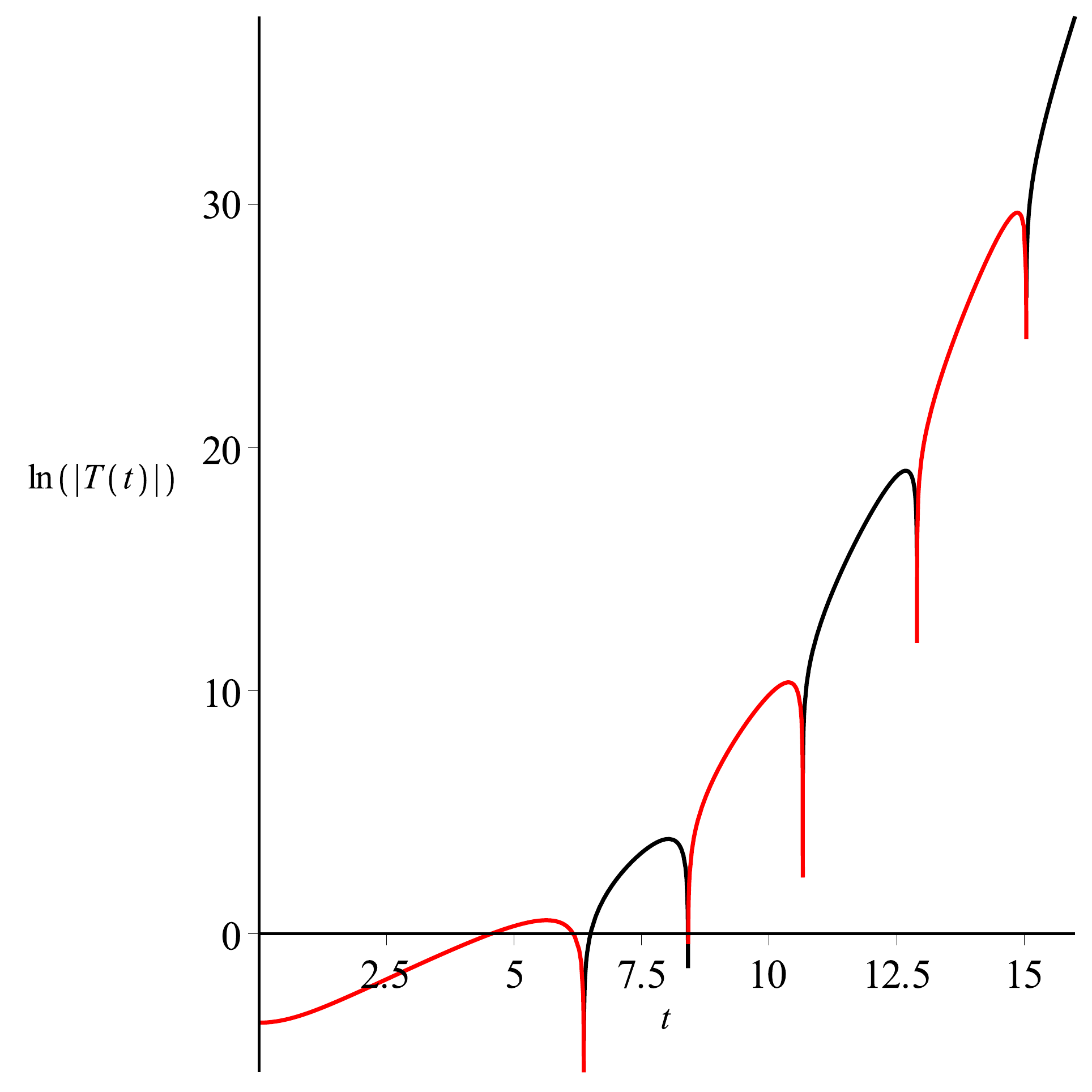}}
	\hfill
	\subfloat[]
	{\includegraphics[width=0.43\textwidth]{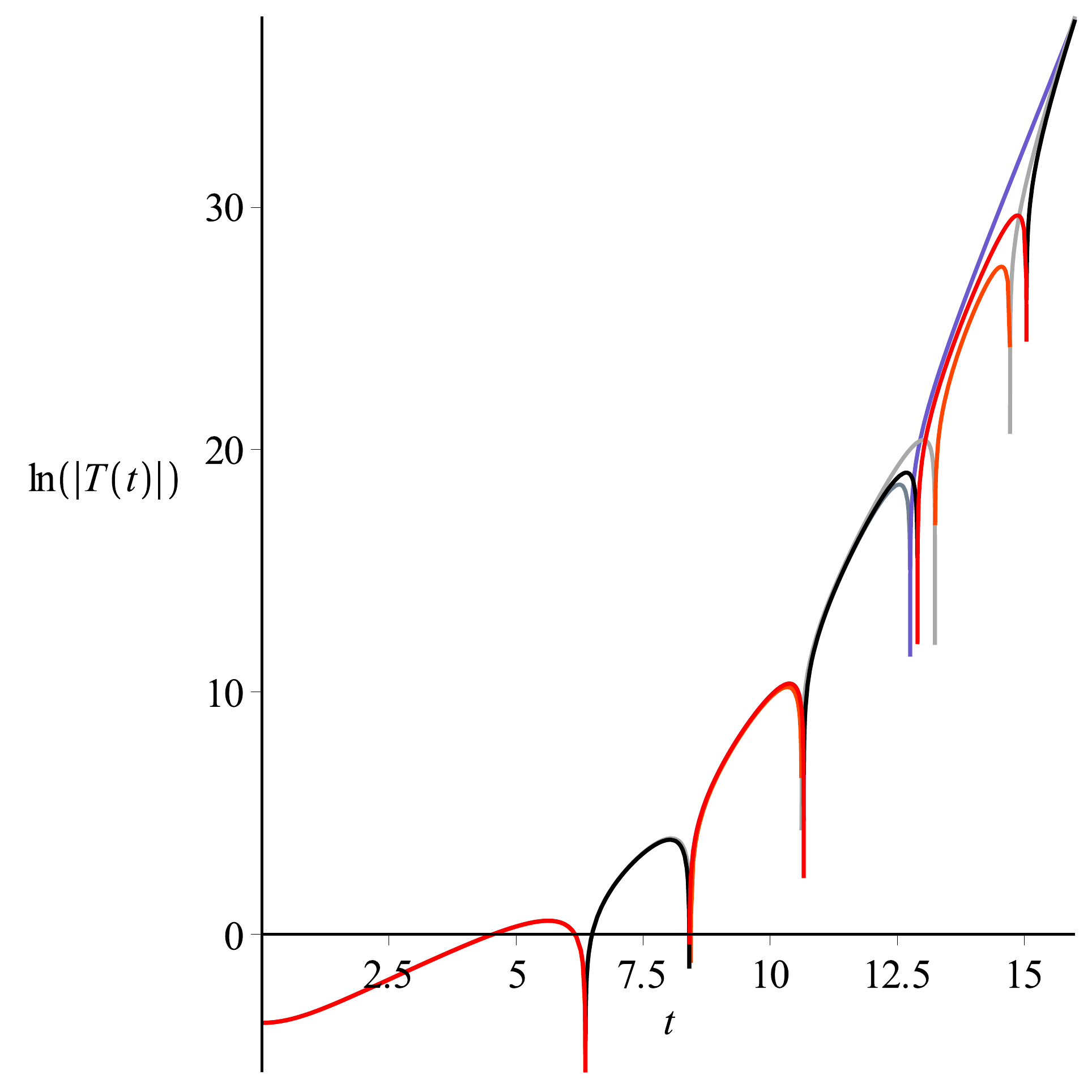}}
	\caption[The tachyon profile with only edge 
	coefficients]{The 
	tachyon profile $T(t)$ with only the $\beta_{n}^{(0)}$ 
	coefficients considered.  This is plotted for $\ln|\lambda|=-4$, 
	where the approximation is valid.  a) Black is for positive values 
	of $T(t)$ and red is for negative values.
	b) The tachyon profile with all \code{Suave} coefficients from table 
	\ref{tab.rtc.solutionmodesS} is also shown in orange and grey, 
	and where the \code{Cuhre}-only results deviate is indicated with a 
	blue line.  We can see that the \code{Suave} results are 
	qualitatively equivalent to the ones we use.}\label{fig.rtc.edge-profile}
\end{figure}

We begin our analysis with the case $|\lambda|\ll1$, where only the 
coefficients $\beta_{n}^{(0)}$ need to be 
considered.  Following the notation of 
\cite{Kiermaier:2007ba,Kiermaier:2010cf}, we will 
refer to these coefficients as 
$\beta_{n}\eqdef\beta_{n}^{(0)}$.
We will focus on \eq{eq.rtc.Tx-ln}, which receives significant 
contributions from the first term in parentheses when $t>0$ and from 
the second term when $t<0$.  Knowing that $T(t)$ is an even 
function, we will assume $t>0$ and not need to consider the second 
term.  Since we are considering $-1\ll\lambda<0$, each term in the 
sum of \eq{eq.rtc.Tx-ln} will be suppressed by the exponential until 
$t$ is large compared to $-\ln|\lambda|$.  For a large fixed $t$, 
terms with $j>0$ will be small relative to others, so only the $j=0$ 
coefficients need to be considered.  Since this is the case, the 
tachyon profile does not depend on the 
renormalization constants $C_{0}$ and $C_{1}$ at all.  This had to be 
the case since there is no renormalization when all of the marginal 
operators have momentum in the same direction.  We can then 
unambiguously plot the tachyon profile for small $|\lambda|$.  In 
figure \ref{fig.rtc.edge-profile} we see $\ln|T(t)|$ for 
$\ln|\lambda|=-4$.  \comment{Only half is shown since $T(t)$ is a symmetric 
function.}  Each ``peak'' represents the range of $t$ for which $T(t)$ 
is dominated by a specific exponential in the sum.  For different 
values of $|\lambda|$ the shape of the oscillating part of the 
tachyon profile remains unchanged, and the whole half-plot shifts 
horizontally, with the size of the plateau in the middle changing as 
expected.  

\begin{figure}
	\subfloat[]
	{\label{fig.rtc.tachyon-zero-edge}\includegraphics[width=0.43\textwidth]{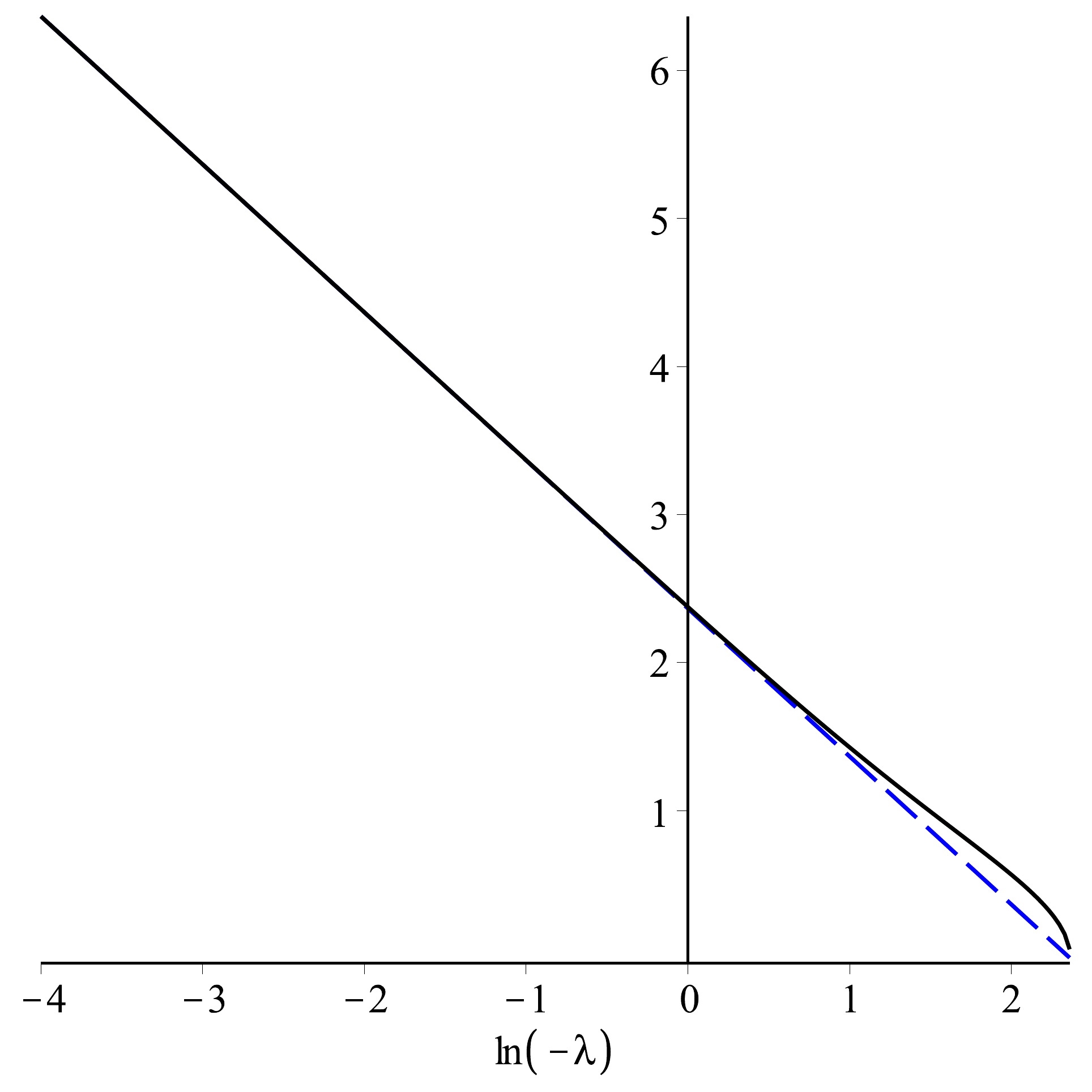}}
	\hfill
	\subfloat[]
	{\label{fig.rtc.tachyon-zero}\includegraphics[width=0.43\textwidth]{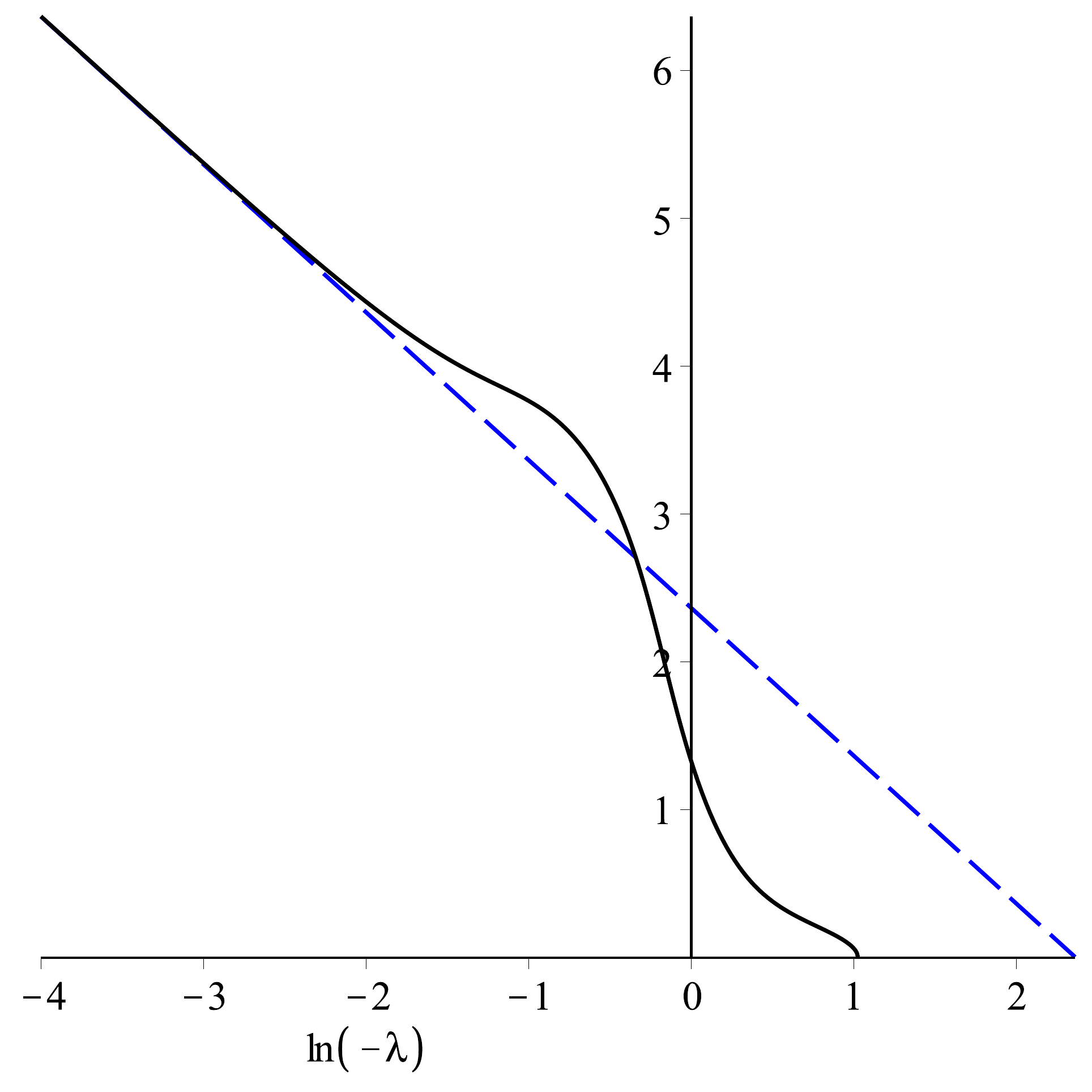}}
	\caption[Location of the first zero of the rolling tachyon profile]
	{The time of the first zero of the tachyon as a 
	function of $\ln|\lambda|$.  An approximation to the asymptotic 
	behaviour is shown as a dashed line.  a) Only the $\beta_{n}^{(0)}$ coefficients 
	are considered, so the plot is not valid for large $|\lambda|$.  
	b) The whole tachyon profile is considered with coefficients from 
	the fit of figure \ref{fig.rtc.opt-RC0kp34}.}
	\label{fig.rtc.tachyon-zeros}
\end{figure}

\begin{figure} \centering
	\subfloat[]
	{\label{fig.rtc.edge-gamma-slope}\includegraphics[width=0.5\textwidth]{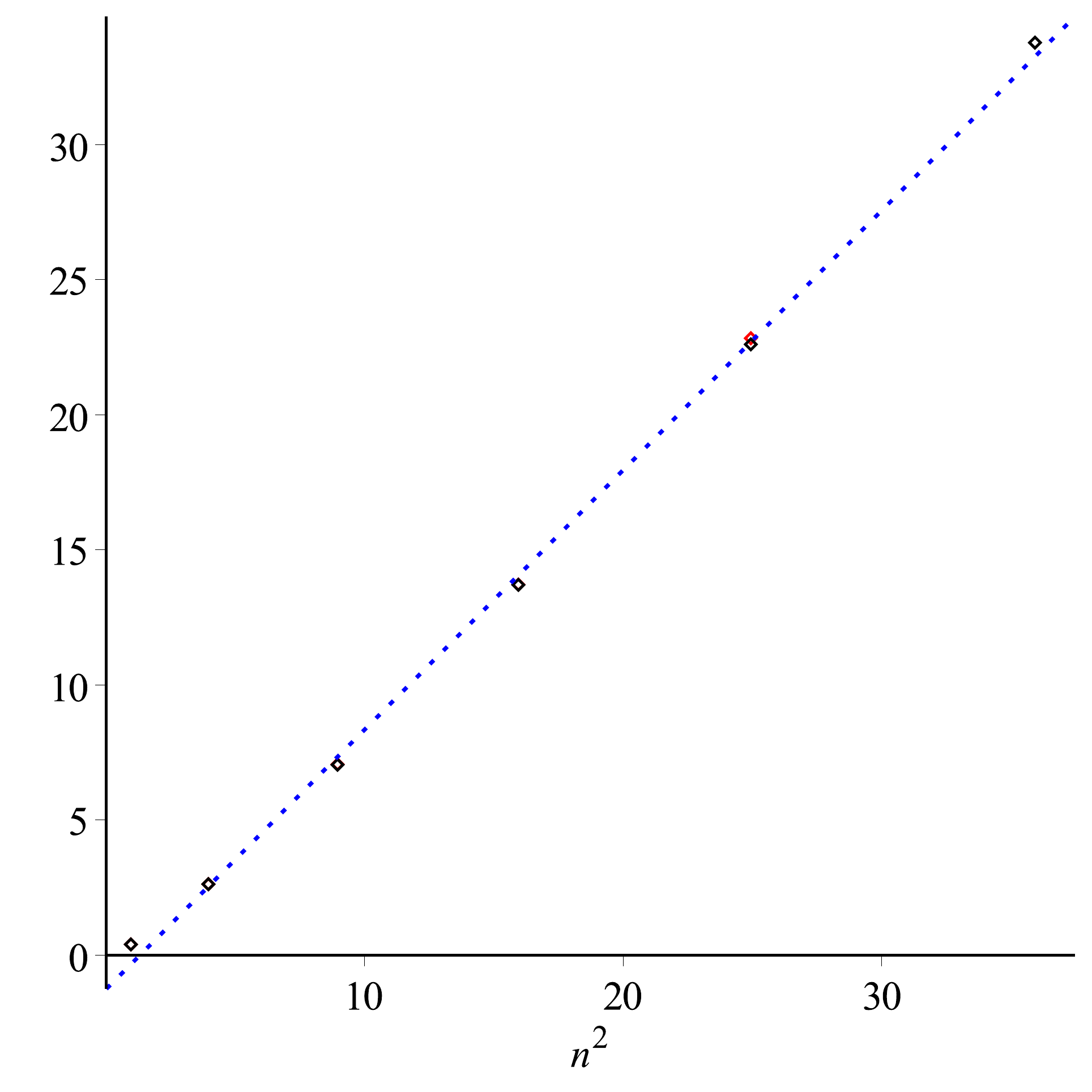}}
	\subfloat[]
	{\label{fig.rtc.edge-gamma-all}\includegraphics[width=0.5\textwidth]{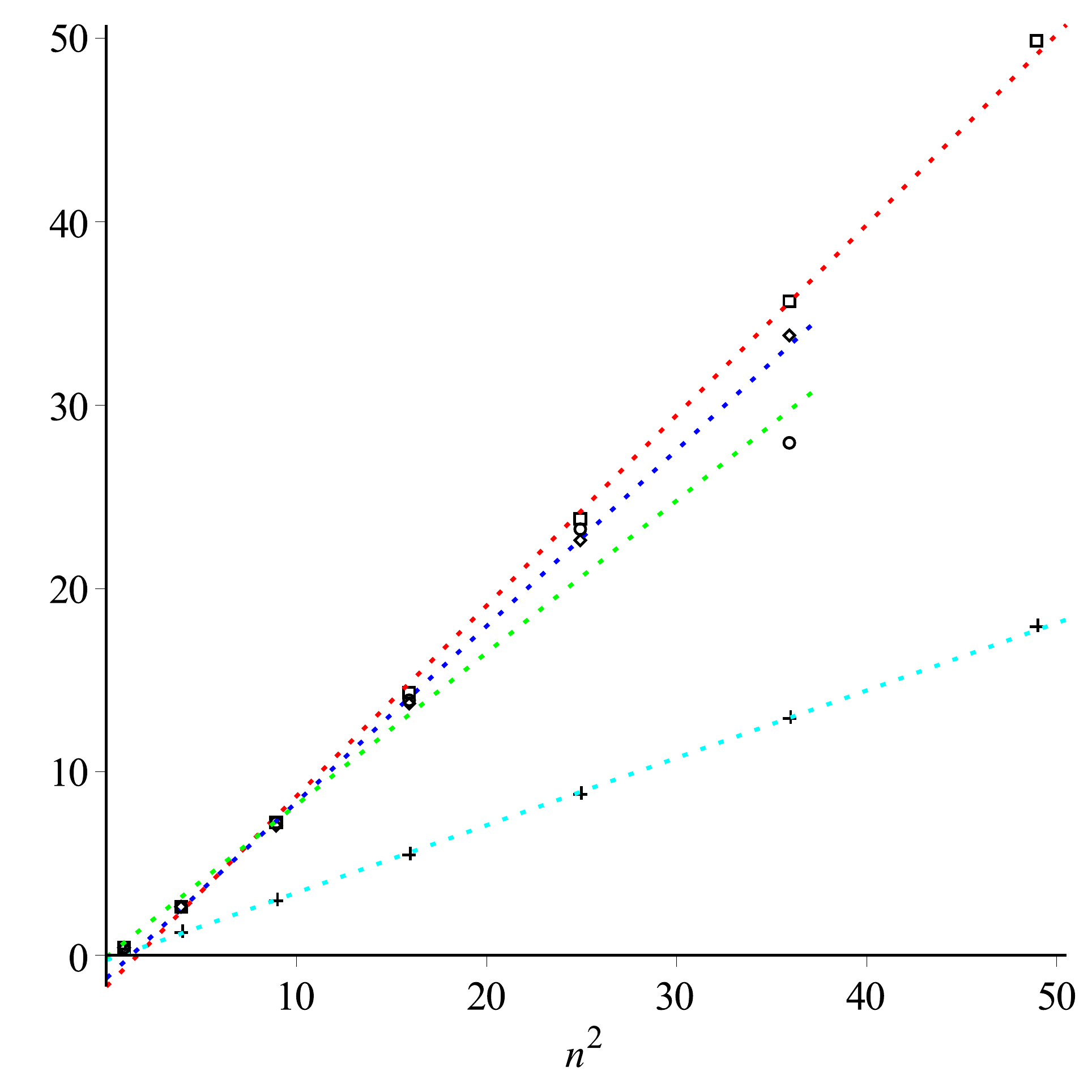}}
	\caption[Linearized trend for the $j=0$ coefficients in the 
	singular rolling tachyon profile based on our computations]
	{The falloff of the edge 
	coefficients shown as $-\ln\beta_{n}^{(0)}$ versus $n^{2}$.  A 
	linear graph indicates that $\beta_{n}\propto e^{-\gamma n^{2}}$ 
	holds, with $\gamma$ given by the slope.  The best linear fit is 
	also shown.  a) The solution shown here, with slope $0.9599$.  
	Red points are \code{Suave} values. 
	b) Our solution as well as the other three presented in table 
	\ref{tab.rtc.korzmodes}.  Square points are for 
	\cite{Kiermaier:2007ba}, crosses are for \cite{Kiermaier:2010cf}, 
	and circles for \cite{Coletti:2005zj}.}
	\label{fig.rtc.edge-gamma}
\end{figure}

The size of the plateau which describes the time when the D-brane 
exists can be estimated by the time of the first zero of the tachyon 
profile.  This time is plotted in figure 
\ref{fig.rtc.tachyon-zero-edge}, 
and is linear for the region where $|\lambda|\ll1$ is 
valid.  The slope is $-1$ as it had to be from \eq{eq.rtc.Tx-ln} when 
$t$ is significantly larger than $0$.  
We can also examine the ``period'' of the oscillations.  The 
oscillations result from each exponential overtaking the one before, 
so we can calculate an estimate of their spacing by setting adjacent 
terms to be equal.  
\begin{subequations}\ba
\beta_{n}e^{n(\ln|\lambda|+t_{n})}&=\beta_{n+1}e^{(n+1)(\ln|\lambda|+t_{n})}\\
t_{n}&=-\ln|\lambda|+\ln\left(\frac{\beta_{n}}{\beta_{n+1}}\right)\\
\Delta t_{n}&=\ln\left(\frac{(\beta_{n})^{2}}{\beta_{n-1}\beta_{n+1}}\right)\label{eq.rtc.oscspacing}
\end{align}\end{subequations}
While there is no reason to expect this a priori, let us suppose that 
$\Delta t$ is a constant.  In this case we have 
\begin{subequations}\be
\frac{\beta_{n+1}}{\beta_{n}}=e^{-\Delta t}\frac{\beta_{n}}{\beta_{n-1}}~,
\ee
which is a recursion relation with the solution 
\be \label{eq.rtc.edge-coeff-hypothesis}
\beta_{n}\propto \rho^{n}e^{-\frac{n^{2}}{2}\Delta t}~.
\ee\end{subequations}
The factor $\rho^{n}$ can always be removed by taking 
$\beta_{n}^{(j)}\rightarrow\frac{\beta_{n}^{(j)}}{\rho^{n}}$ and 
simultaneously $\lambda\rightarrow\lambda\rho$, which does not alter 
the tachyon profile.
In one particular solution for the rolling tachyon with regular OPE, 
Kiermaier, Okawa, and Soler \cite{Kiermaier:2010cf} found that their 
solution's coefficients had the asymptotic behaviour 
\be
\beta_{n}\sim e^{-\gamma n^{2}+O(n\ln n)}~,
\ee
and in \cite{Schnabl:2007az} it was shown that a solution equivalent 
to the one in \cite{Kiermaier:2007ba} has coefficients which closely 
fit $b_{n}\sim e^{-\gamma n^{2}}$ without significant corrections.  
We have just shown that this same recurring pattern can be derived 
from the assumption of exponentially growing oscillations with 
constant period.  In figure 
\ref{fig.rtc.edge-gamma} we see the 
best fit lines for our $j=0$ coefficients, as well as those of 
several other known solutions, to the form $\beta_{n}\sim e^{-\gamma n^{2}}$.  
This was only predicted to be a fit for one solution at large $n$, but 
we see good agreement in all cases, even with $n$ never rising past 6 
or 7 for any of the solutions considered.  In figure 
\ref{fig.rtc.edge-gamma-slope} the fit is to the deterministic 
results of table \ref{tab.rtc.solutionmodesRC} for $n\leq5$ and the 
\code{Suave} result for $n=6$, but the \code{Suave} results with 
smaller $n$ are shown as the red points for reference. 
The \code{Cuhre} value for $\beta_{6}^{(0)}$ cannot be shown on a 
logarithmic plot since it has the wrong sign.  It is curious that the 
coefficients fall so close to the $e^{-\gamma n^{2}}$ lines without 
any correction, even such as choosing $\rho\neq1$ in 
\eq{eq.rtc.edge-coeff-hypothesis}.
While this trend was derived in our case from a constant 
period of oscillation, if it holds at higher $n$ it guarantees that 
the edge coefficients are a convergent series.  The fact that all of 
the solutions appear to behave similarly suggests that they are also 
all convergent.

\subsection{Large $\lambda$}

Once we loosen the $|\lambda|\ll1$ restriction, we must consider 
all of the coefficients $\beta_{n}^{(j)}$ and search for patterns 
there.  Due to the small number of coefficients, and particularly the 
small number of rows with constant $j$, it is not possible to 
get a good understanding of any patterns or asymptotics for these 
coefficients, but we can speculate as to possible trends.  
The first thing we notice from table 
\ref{tab.rtc.solutionmodesRC} is that the sign of the coefficients 
appears to alternate as $(-1)^{\frac{j}{2}}$.  This is not strictly 
true even for the coefficients we have calculated, however, as 
choosing non-zero $C_{0}$ and especially $C_{1}$ will alter many of 
the coefficients and can affect their sign.  With only a small number 
of the coefficients known, we do not know whether the large $n$ 
asymptotics are fixed or can be changed by a choice of the two free 
parameters.  We can, however, attempt to force a few patterns and see 
which appear more naturally.

\begin{figure}
	\subfloat[]
 	{\label{fig.rtc.opt-RC0kp34}\includegraphics[width=0.32\textwidth]{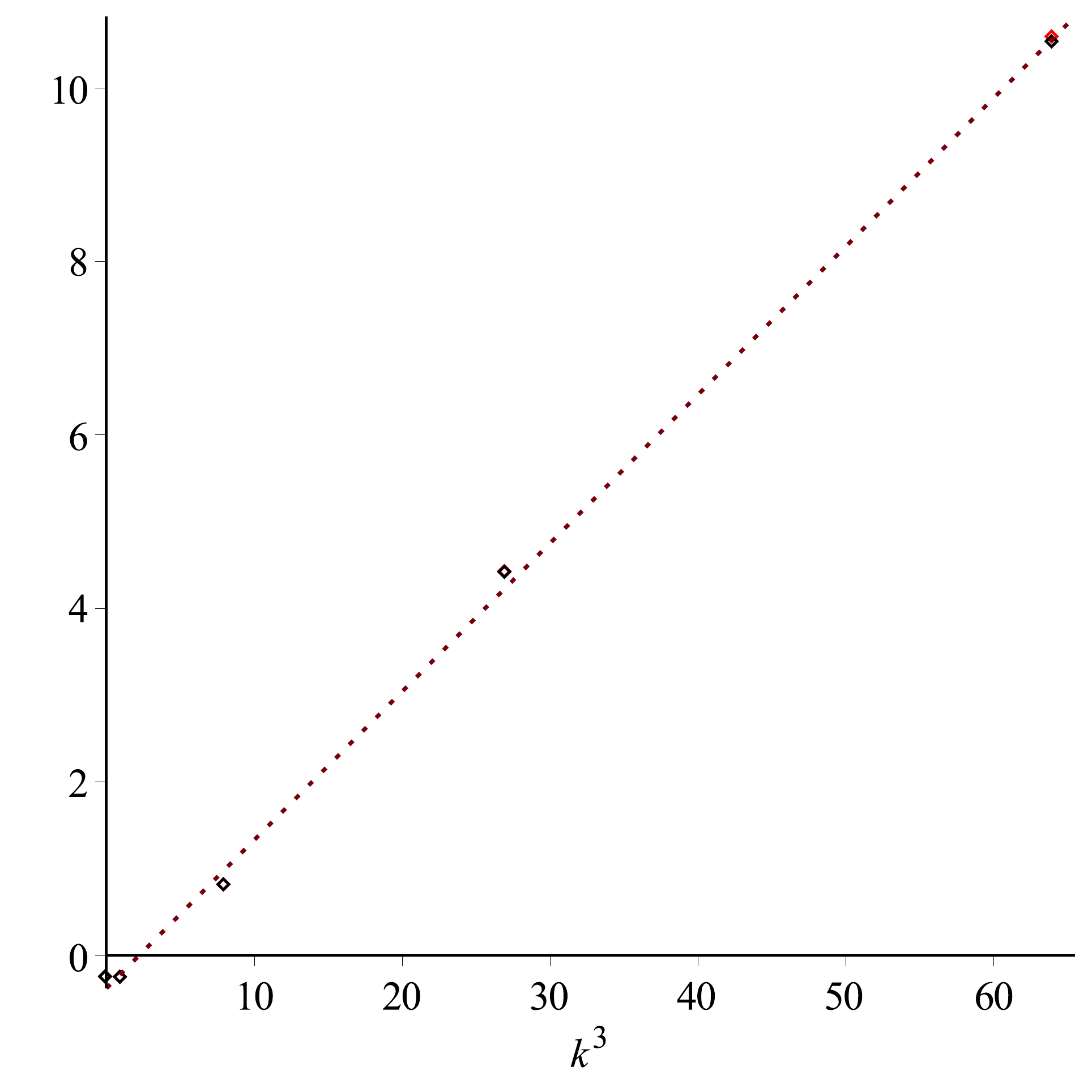}
 		\includegraphics[width=0.32\textwidth]{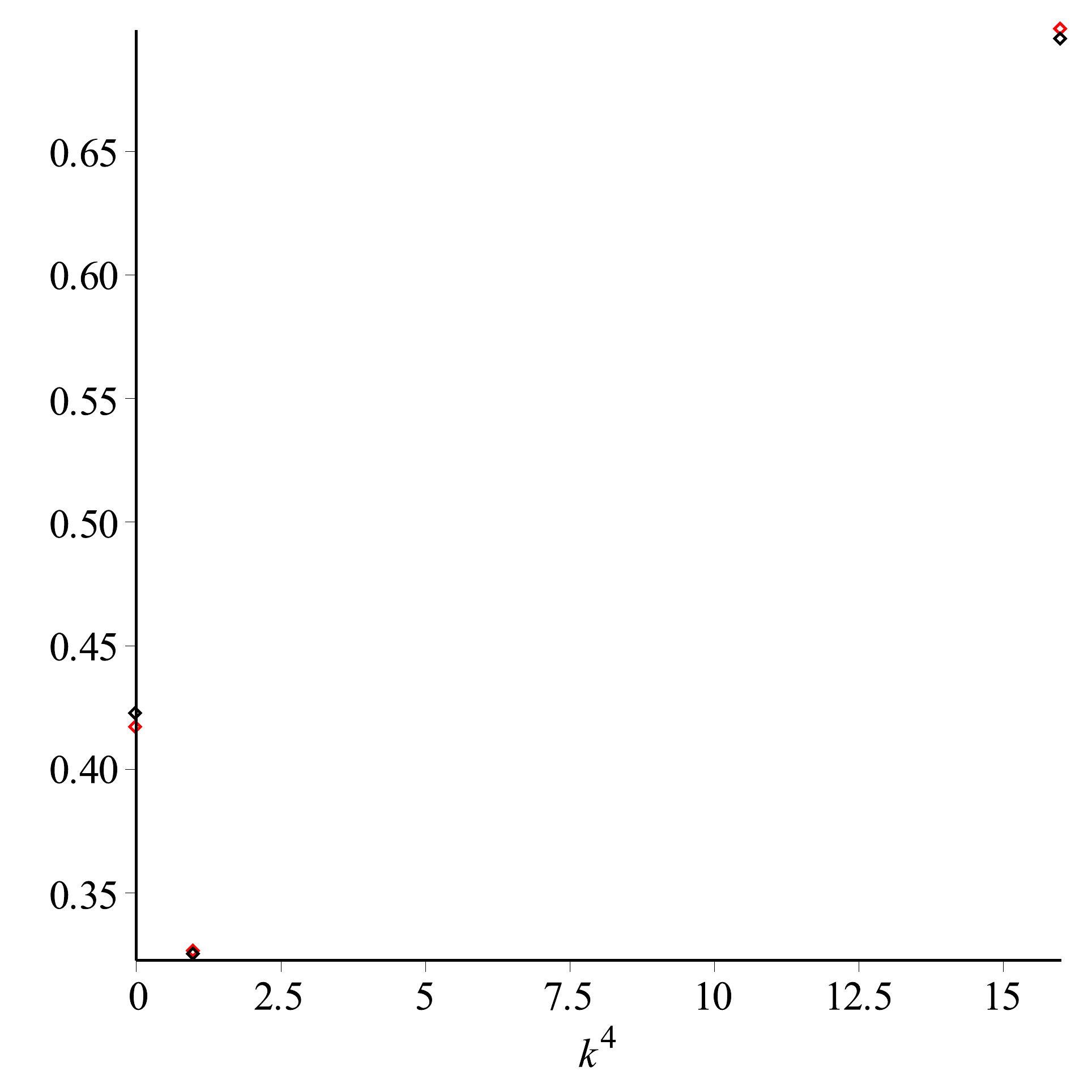}}
	\subfloat[]
	{\label{fig.rtc.opt-slopeonlykp3}\includegraphics[width=0.32\textwidth]{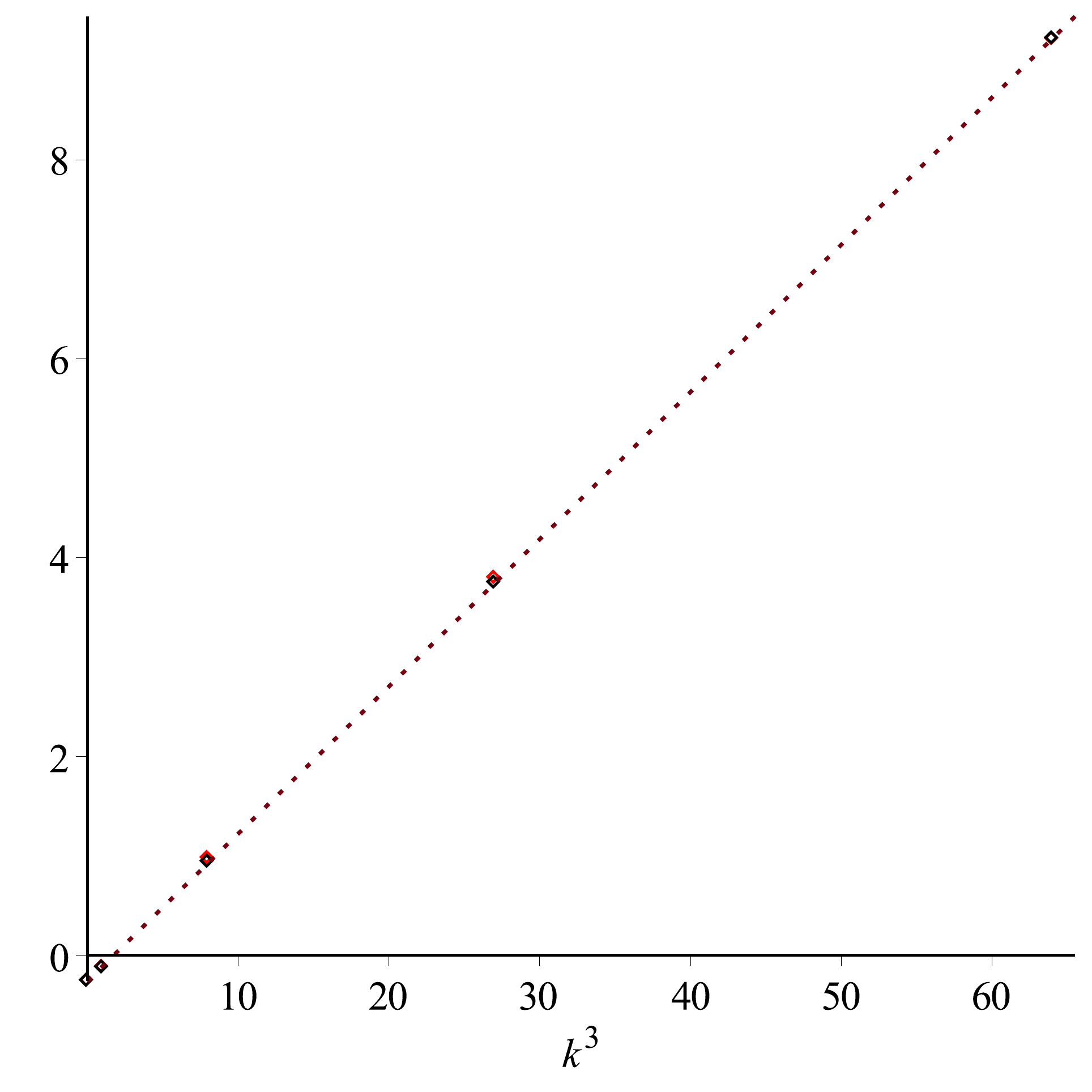}}\\
	\subfloat[]
 	{\label{fig.rtc.opt-slopeonlykp2}\includegraphics[width=0.32\textwidth]{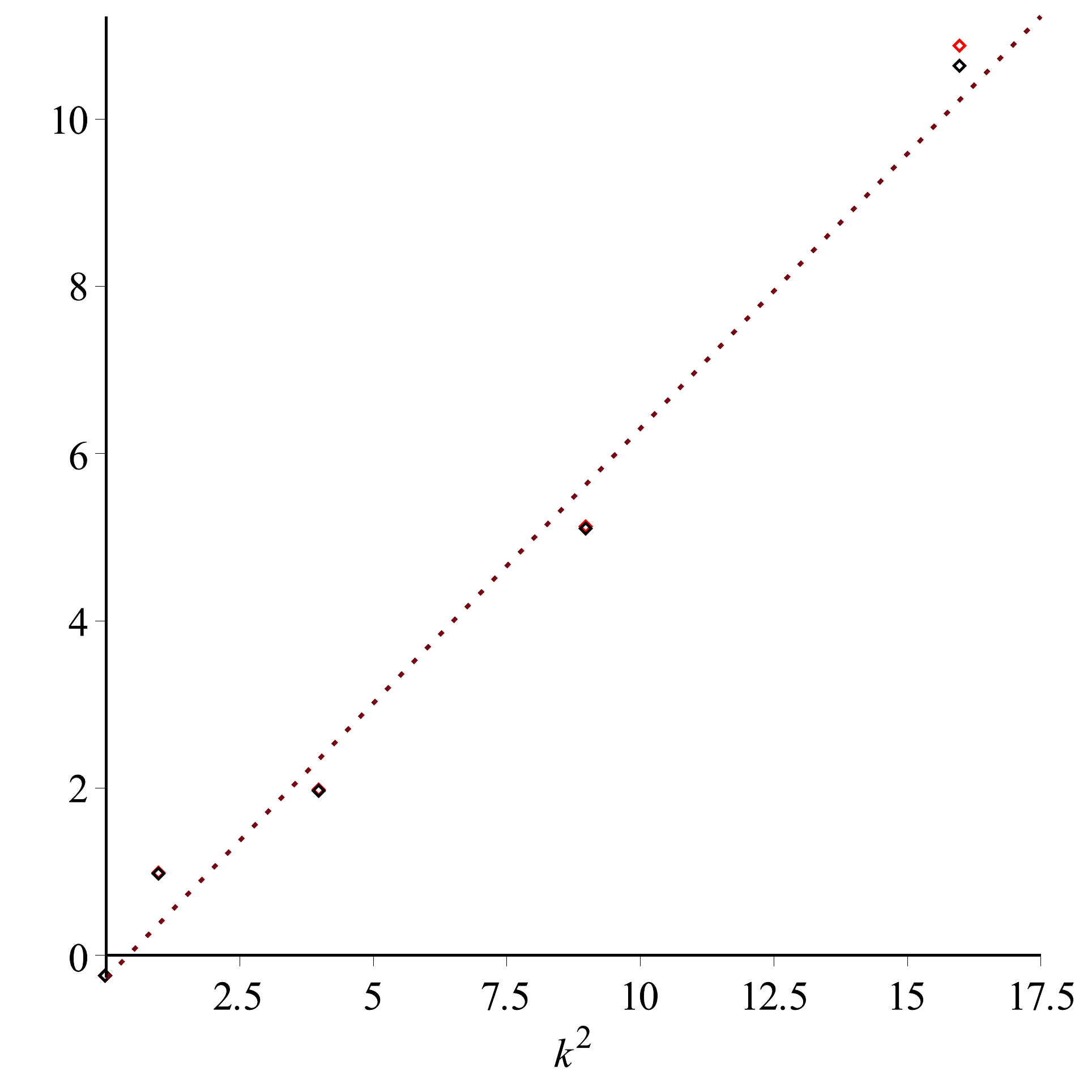}}
 	\subfloat[]
 	{\label{fig.rtc.opt-slopeallkp2}\includegraphics[width=0.32\textwidth]{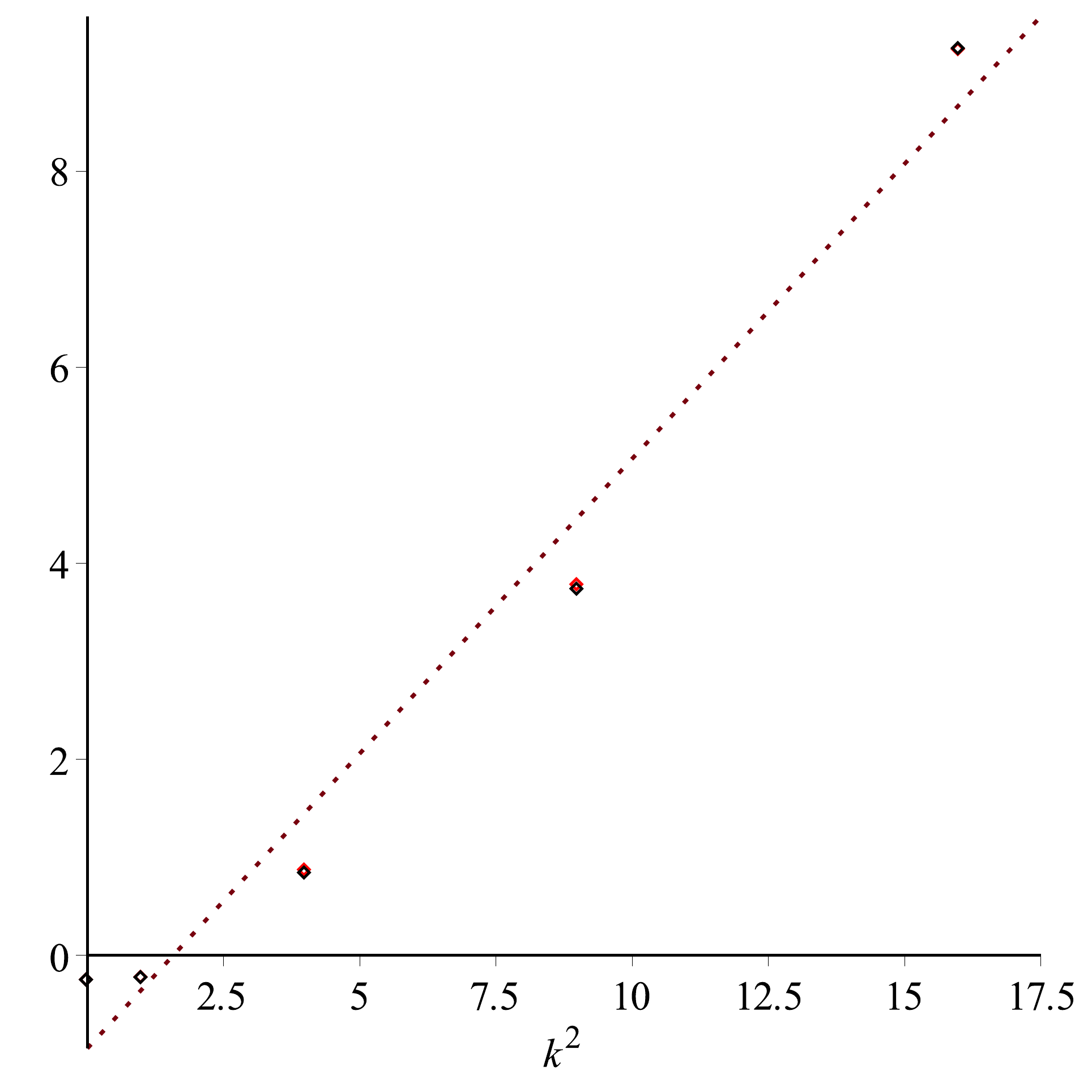}
 	\includegraphics[width=0.32\textwidth]{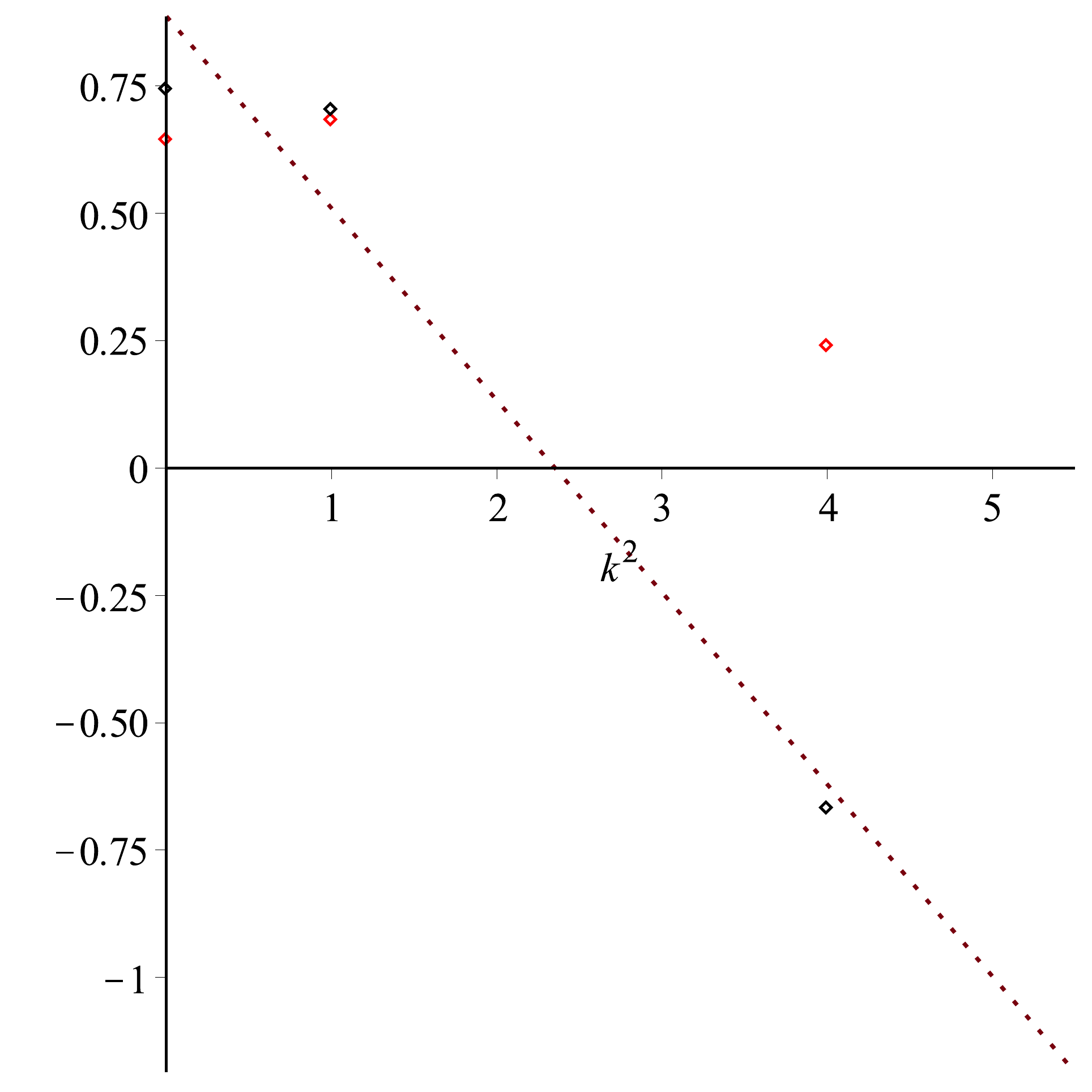}} \\
	\caption[Plots showing several attempts to determine a trend for 
	the bulk coefficients in our rolling tachyon computations]
	{$\ln|\beta_{n}^{(j)}|$ plotted vs. functions of $k=n-2j$ with 
	several different choices for $C_{0}$ and $C_{1}$. Black 
	points are \code{Cuhre} values while red points are from the 
	\code{Suave} algorithm. 
	In a) we set $C_{0}=C_{1}=0$ and plot $j=1$ coefficients on the 
	left and $j=2$ on the right. 
	The $j=1$ coefficients with $C_{0}$ and $C_{1}$ optimized for the 
	best linear fit appear in b), and 
	c) attempts the same linear fit assuming a $k^{2}$ horizontal 
	axis rather than $k^{3}$. 
	d) attempts a linear fit to both the $j=1$ and $j=2$ sets of 
	coefficients assuming a $k^{2}$ horizontal axis, with $j=1$ 
	coefficients on the left and $j=2$ on the right.}
	\label{fig.rtc.opts}
\end{figure}

As a first choice we pick $C_{0}=C_{1}=0$ and notice that the $j=1$ 
coefficients appear to be a good fit to $\beta_{n}^{(1)}\sim 
e^{-\gamma_{1}k^{3}}$ with $k=n-2j$, which is shown in figure 
\ref{fig.rtc.opt-RC0kp34}.  While this can be made to fit even better 
by a choice of renormalization constants, this would lead to some of 
$\beta_{n}^{(2)}$ being less than zero or to that row having 
increasing magnitudes.  On the other hand, if we attempt to pick 
renormalization constants which are a fit to $\beta_{n}^{(1)}\sim 
e^{\gamma_{1}k^{2}}$, as shown in figure 
\ref{fig.rtc.opt-slopeonlykp2}, we do not find as good a fit.  The 
same is true of $\beta_{n}^{(1)}\sim e^{\gamma_{1}n^{2}}$ using $n$ 
instead of $k$ in the exponent.
It appears that the $j=1$ coefficients have a tendency towards the cubic 
exponential decay, while for $j=2$ we lack enough points to 
reach any conclusions.
The red points in figure \ref{fig.rtc.opts} again represent the 
\code{Suave} coefficients, and we see that $\beta_{n}^{(2)}$ have 
significantly different values once the renormalization constants are 
changed, but a look at table \ref{tab.rtc.solutionmodesS} suggests 
that this is mainly due to large errors in the \code{Suave} 
coefficients, so it is unlikely that the deterministic plots would 
change significantly if more sample points were used.

While five points is not a lot of data, the 
$\beta_{n}^{(1)}$ coefficients suggest that each row with constant 
$j$ may eventually be a convergent series for at least some choice of 
renormalization constants.  Showing that the full tachyon profile 
converges when these rows are added together, however, remains 
impossible until much higher order calculations can be performed.  In 
particular, the large dependence of coefficients such as 
$\beta_{6}^{(3)}$ on $C_{1}$ is troubling since it suggests that if 
that constant is of order 1 then the sequence 
$\beta_{n}^{\left(\frac{n-k}{2}\right)}$ with constant 
$k$ could have increasing magnitudes.  Thinking of 
\eq{eq.rtc.Tx-ksum} as 
\be \label{eq.rtc.Tx-betaeff}
T(t)=2\sum_{k=0}^{\infty}\beta_{\text{eff}}^{(k)}(\lambda)\,\cosh(kt)~,
\ee
the effective coefficients $\beta_{\text{eff}}^{(k)}(\lambda)$ would then be 
defined by series which do not converge for non-zero renormalization 
constants.  Looking at table \ref{tab.rtc.solutionmodesRC}, even with 
vanishing renormalization constants, the magnitudes of the terms 
$\beta_{2j}^{j}$ do not drop off very fast, but with alternating sign 
the series may still converge.

\begin{figure}
	\subfloat[]
	{\label{fig.rtc.tachyon-profile-05}\includegraphics[width=0.45\textwidth]{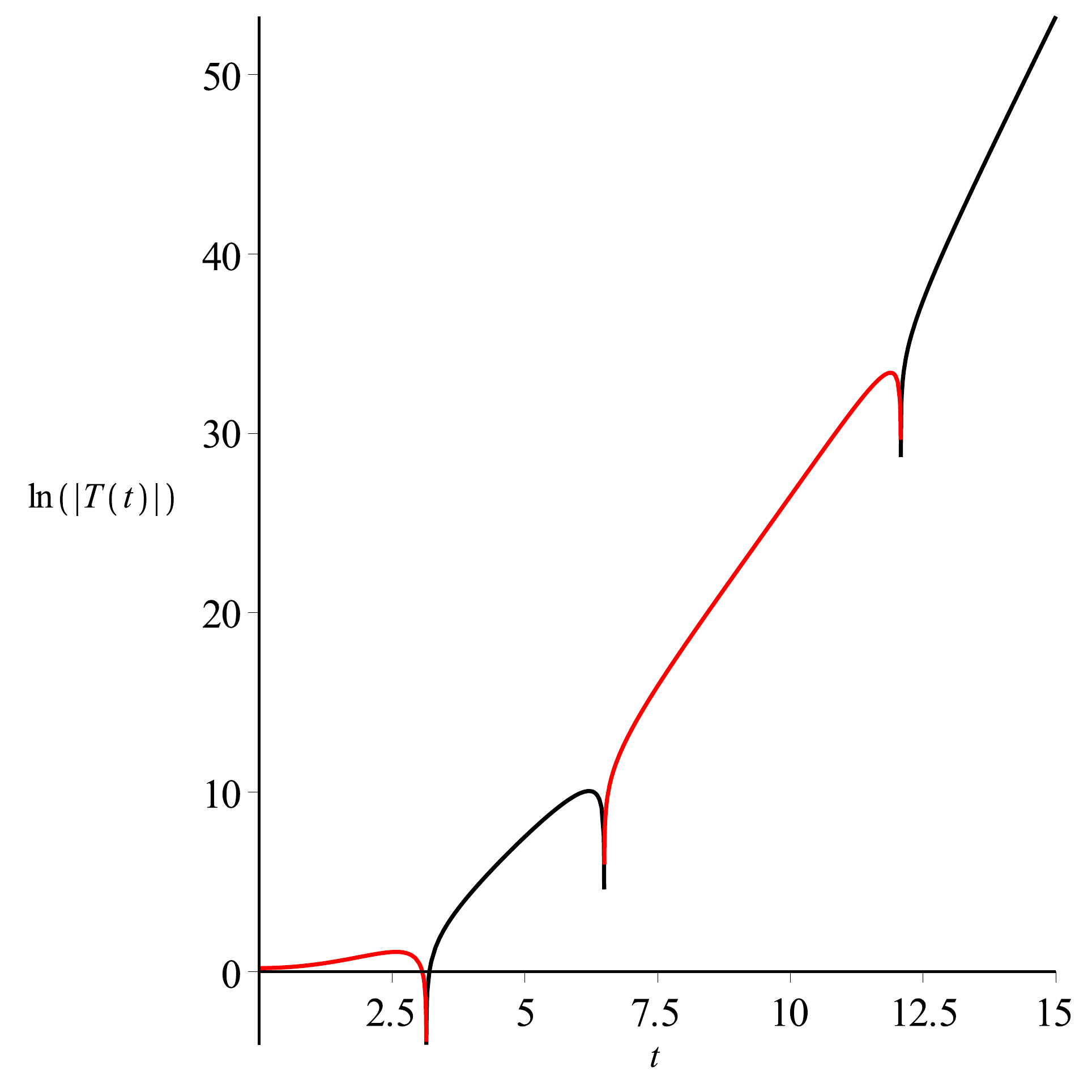}}
	\subfloat[]
	{\label{fig.rtc.tachyon-profile-1625}\includegraphics[width=0.45\textwidth]{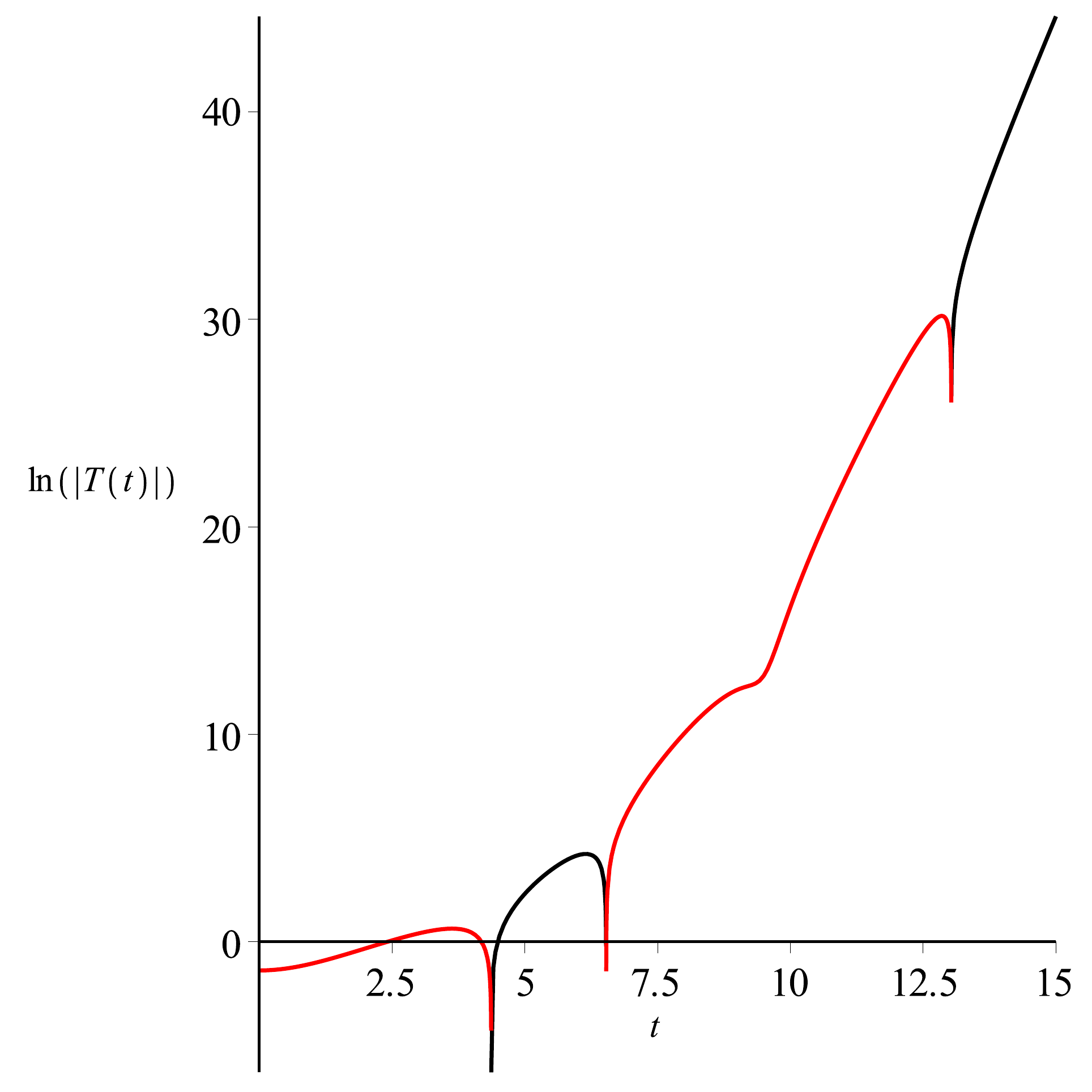}}
	\\
	\subfloat[]
	{\label{fig.rtc.tachyon-profile-17}\includegraphics[width=0.45\textwidth]{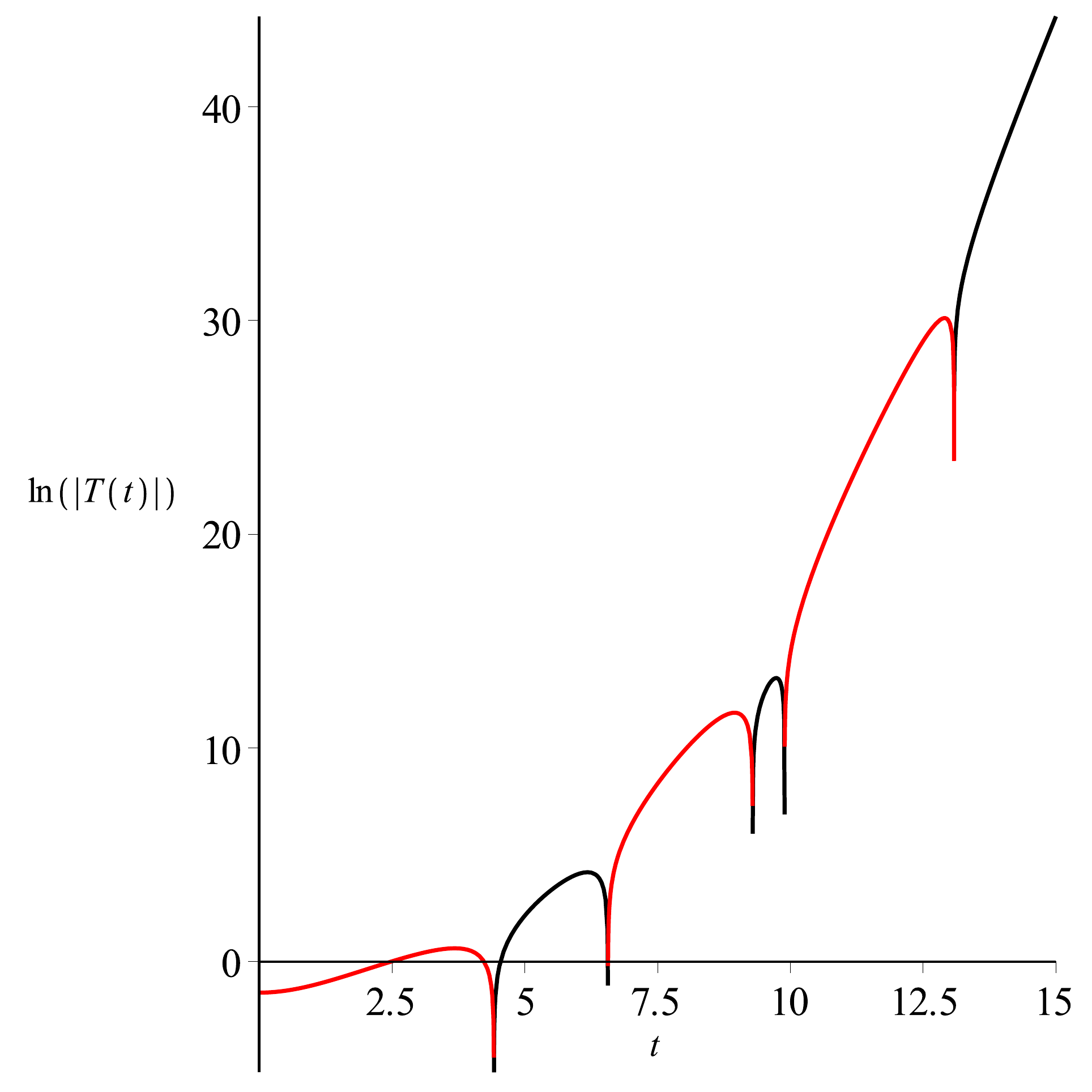}}
	\subfloat[]
	{\label{fig.rtc.tachyon-profile-2}\includegraphics[width=0.45\textwidth]{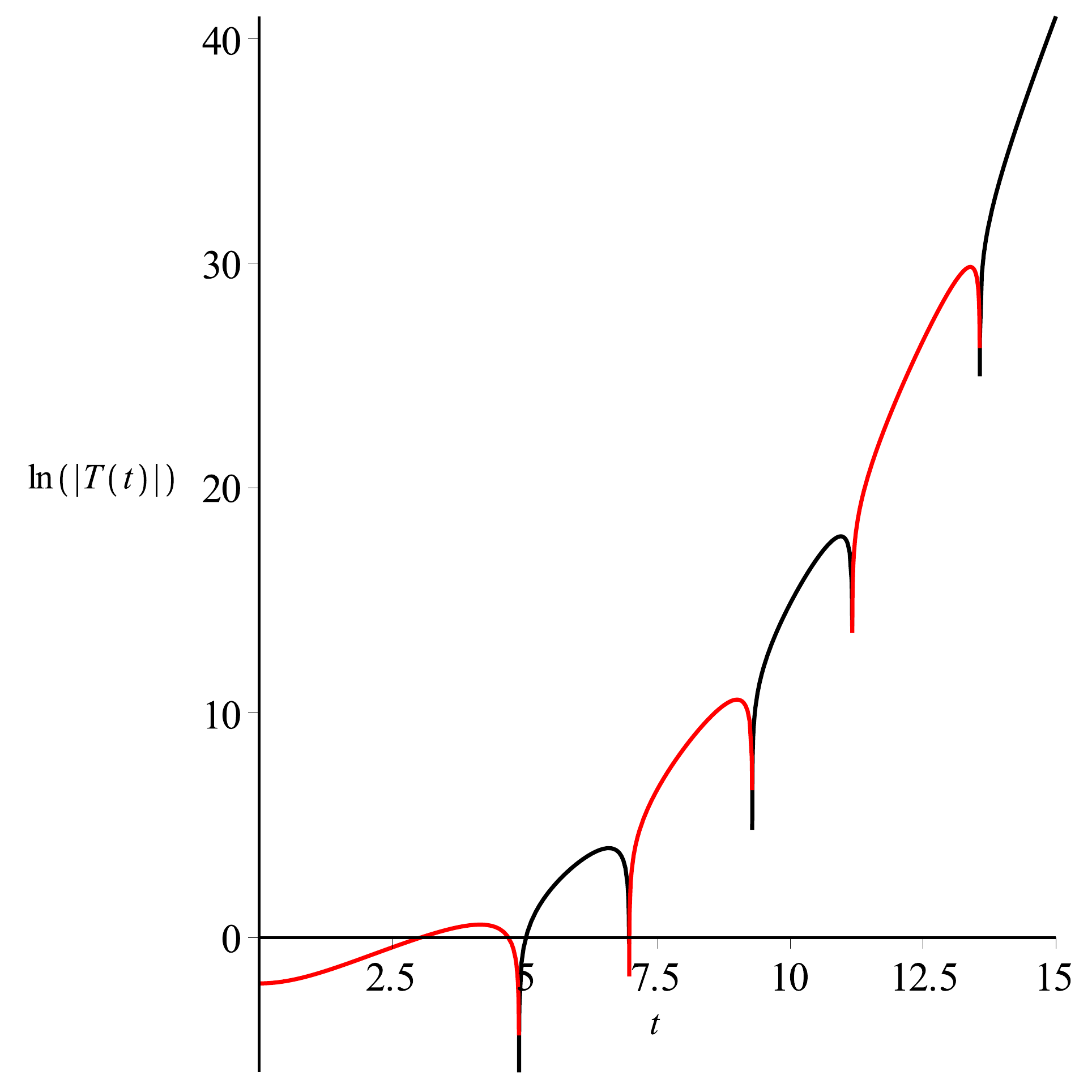}}
	\caption[The tachyon profile produced from our computation of the 
	singular rolling tachyon solution]
	{The log of the tachyon profile at a) $\ln|\lambda|=-0.5$, 
	b) $\ln|\lambda|=-1.92$, c) $\ln|\lambda|=-1.98$,  and d) 
	$\ln|\lambda|=-2.5$.  Positive values represented by black, 
	negative values by red.  The renormalization constants $C_{0}$ 
	and $C_{1}$ are both set to zero.} 
	\label{fig.rtc.tachyon-profile}
\end{figure}

Now that we have seen what the bulk coefficients look like, we can 
begin examining the tachyon profile for larger values of 
$|\lambda|$.  Of course as we do this we must be aware that we are 
missing all coefficients $\beta_{n}^{(j)}$ with $n\geq7$, and as we 
increase the strength of the marginal deformation those coefficients 
will begin to play a larger role, but we can still get a qualitative 
idea of the impact of the bulk coefficients on the tachyon profile.  
Since none of the optimized fits in figure \ref{fig.rtc.opts} were 
significantly better than simply setting the renormalization 
constants to zero, we will choose that from now on.  Other reasonable 
choices will not give results that are qualitatively different.  
For large negative $\lambda$ we see a tachyon profile in figure 
\ref{fig.rtc.tachyon-profile} with fewer 
oscillations than we had with just the edge coefficients from figure 
\ref{fig.rtc.edge-profile}.  As $|\lambda|$ decreases, the additional 
oscillation appears at $\ln|\lambda|\approx-1.948$.  Once 
$\ln|\lambda|\lesssim-2.5$ the profile has stabilized and the plateau 
continues growing as $|\lambda|$ shrinks, just as we know it should 
from our discussion of the tachyon profile for small $|\lambda|$.  
The disappearance of this oscillation for large $|\lambda|$ is because 
the bulk coefficients cannot be neglected in this region, and they 
change the effective coefficients in \eq{eq.rtc.Tx-betaeff}.

The behaviour we see 
for these large values of $|\lambda|$ is not unprecedented; in 
\cite{Kiermaier:2010cf} the tachyon profile had coefficients (seen in 
table \ref{tab.rtc.korzmodes}) which did not decrease as quickly as 
other time-asymmetric solutions.  
That tachyon profile had fewer oscillations for all $\lambda$ because 
some of the exponentials did not dominate for any range of time.  
Aside from the obvious, that the singular OPE case is a symmetric 
function where the D-brane exists for a limited time while with 
regular OPE it exists until it decays at a finite time, the 
qualitative difference between the tachyon profiles seems to be that 
the period and number of oscillations can change this way.  Because 
the strong deformation tachyon profile we have found is similar to 
the profile of \cite{Kiermaier:2010cf}, it suggests that changing the 
strength of the marginal deformation in the time-symmetric case is 
much like changing gauge in the time-asymmetric case.  
If the late time behaviour is equivalent to the tachyon vacuum under 
a time-dependent gauge transformation, as has been hypothesized 
\cite{Coletti:2005zj}, then in this case the gauge transformation 
should depend on both time and the marginal parameter in a 
non-trivial way.  That our solution appears qualitatively like 
time-asymmetric ones for both weak and strong deformation parameter 
suggests that such gauge transformations remain a valid explanation 
of the oscillations in the time-symmetric case.

\section{A Few Technical Details} \label{sec.rtc2.details}
The solution and necessary renormalized operators have been defined 
in \cite{Kiermaier:2007vu} and \cite{Karczmarek:2014wea}, so it is 
possible to numerically compute many quantities up to a reasonable accuracy.  
We use Maple to algebraically manipulate wedge states with insertions 
and produce functions to be integrated.  This means writing routines 
for the star product, BRST operator, and correlation functions, among 
other things.  While the number of terms grows extremely fast, it is 
possible to algebraically construct the solution up to 7th order in 
$\lambda$.  The integrands produced, however, are complicated enough 
that I was only able to compile them for integration up to 6th 
order.  Although the algebraic construction of the wedge states with 
insertions is a lengthy process, it is an exact one which should not 
produce any numerical errors.  Evaluating the resulting integrals, 
however, is done by sampling the integrand at many points and 
estimating a result, and this process unavoidably introduces errors.  

The integration is handled by off-the-shelf C++ routines.  The CUBA 
library appears to be a good choice 
in most cases \cite{Hahn:2004fe}.\footnote{The CUBA library is 
distributed from \url{http://www.feynarts.de/cuba/}.}  It is a collection of four 
algorithms for multi-dimensional numerical integration, three of which 
use pseudo-random sampling while the fourth is a deterministic 
algorithm.  Since we are working at sixth order in $\lambda$ and the 
solution has ghost number one (corresponding to the number of fixed 
moduli), there are never more than five integrated 
coordinates in a given integral.  While Monte Carlo algorithms do 
scale better as the dimension rises, in five or less dimensions it 
appears that the deterministic algorithm, \code{Cuhre}, is slightly 
more efficient.  As we will see, \code{Cuhre} is also more reliable 
in most cases.  Unfortunately, it only integrates 
functions of more than one variable, so in the one-dimensional case 
we use the \code{QAG} 
routine from the GNU scientific library.\footnote{The GNU scientific 
library is found at \url{https://www.gnu.org/software/gsl/}.}  
Each of the routines in the CUBA and GNU libraries 
provides its own error estimate, and the CUBA library routines also 
provide a chi-square estimate of the probability that the error is 
sufficient.  

A single quantity to be calculated numerically, such as an individual 
term in the tachyon profile of table \ref{tab.rtc.solutionmodesRC} or 
in one of the consistency checks of tables 
\ref{tab.rtc.checks1} and \ref{tab.rtc.checks2}, generally consists of 
a small number of integrals.  Each integral contains all of the terms 
in the solution which are integrated over a given number of 
coordinates, or equivalently all of the terms with the same number of 
integrated operators.  Most quantities are the sum of two integrals, 
but a few are only a single integral, and quantities in the action 
can consist of more than two.

\subsection{Convergence}

In order to get as much data as possible, the collection of integrals 
we look at here will include all of the ones used in calculating 
the tachyon profile as well as the action and several components of 
the equation of motion.  The action and equation of motion will be 
discussed as checks of consistency in section \ref{sec.rtc.checks}.

It is difficult to study convergence of the one-dimensional integrals 
due to the fact that the \code{QAG} algorithm does not report the 
number of samples used.  We can, however, take several full sets of 
data for these integrals and compare the different calculations of 
the same integrals.  We find that they all agree with each other well 
within the error estimates.  The only troubling one-dimensional 
case is that of a constant integrand.  This can be seen in the kinetic 
energy of the solution at fourth 
order in $\lambda$, which is the particularly simple quantity 
$\int_{0}^{1}dt~\frac{3}{2}-\frac{3}{2}$.  The integration algorithm 
performs operations on the constant causing tiny roundoff errors 
which, when the constant value is subtracted, causes the result to 
differ from zero.  This would not be a problem except that error 
analysis in numerical integration is based on variation of the 
integrand, and as such gives an estimated error which is extremely 
small.  While in principle error estimates should account for the 
roundoffs inherent in their algorithm, in practice this does not seem 
to be the case.  The error estimate becomes small enough that the 
reported result can actually be incorrect, and even unstable with 
respect to changing the desired accuracy.  Most integrands worth 
using a numerical algorithm to 
integrate undoubtedly vary enough that this is not normally an 
issue.  This is the only instance of such a problem that we 
encounter, but since the integral is trivial we do not need to rely 
on numerical integration for its value.  

\begin{figure}[t]
	\subfloat[]
	{\label{fig.rtc2.reserr-C}\includegraphics[width=0.45\textwidth]{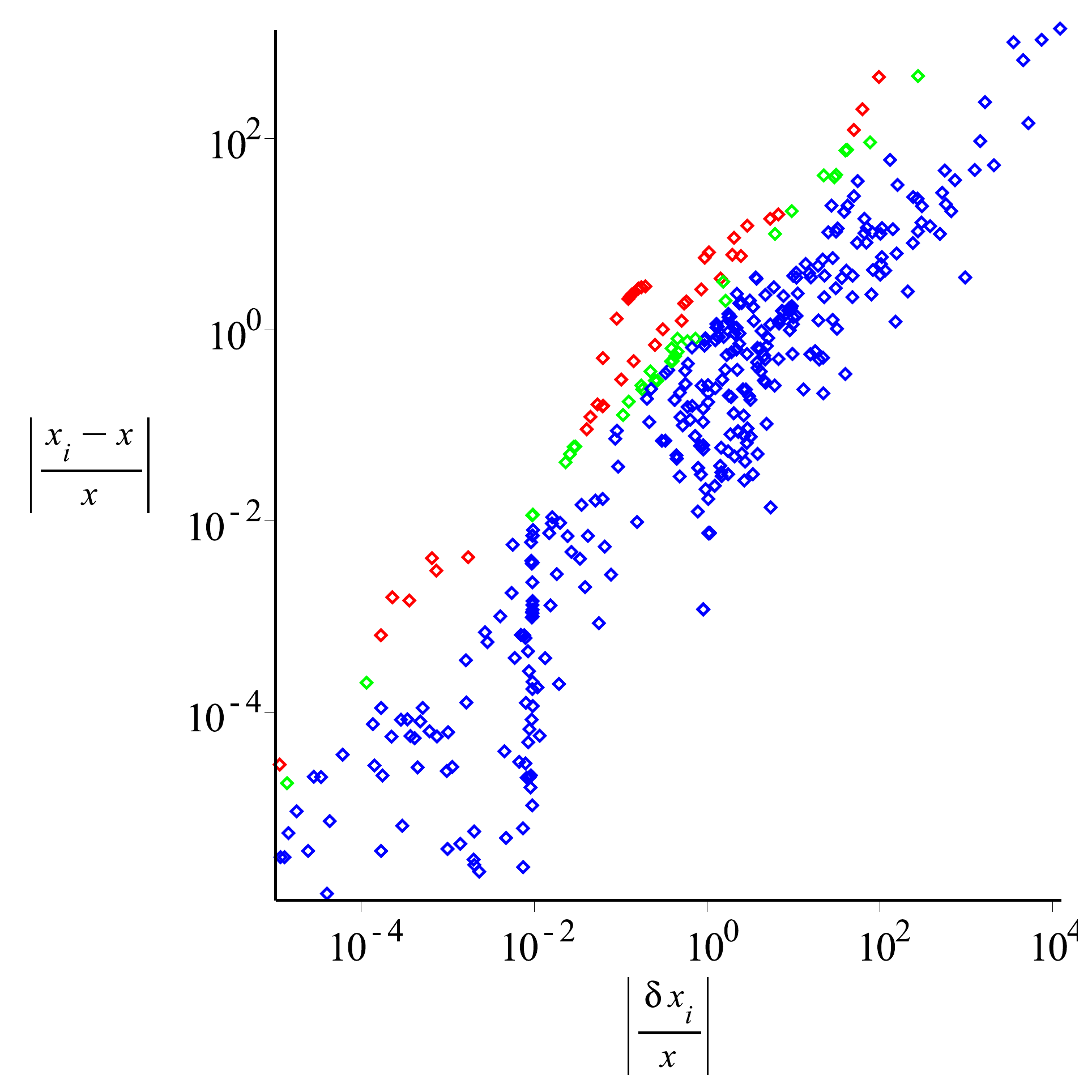}}
	\hfill
	\subfloat[]
	{\label{fig.rtc2.reserr-S}\includegraphics[width=0.45\textwidth]{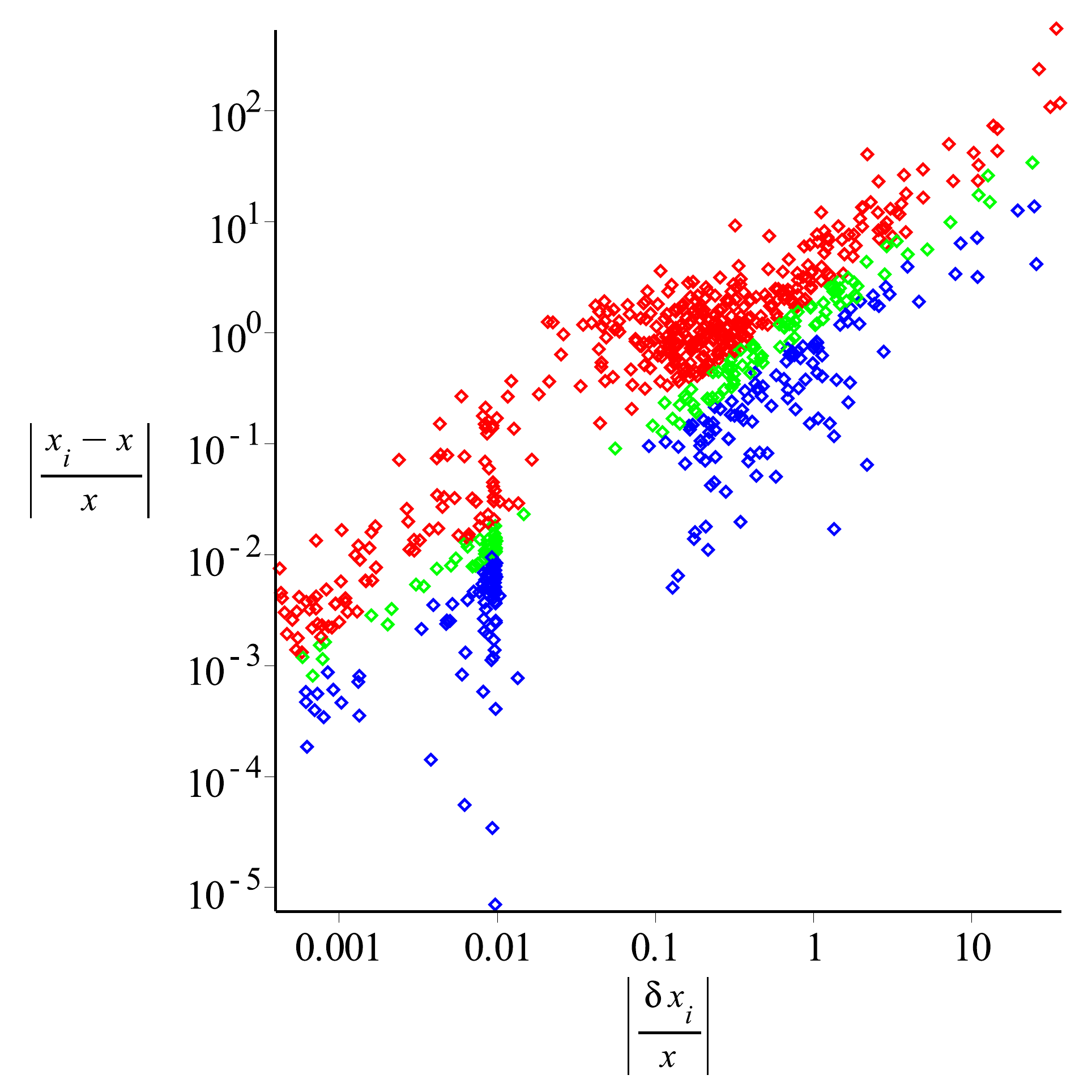}}
	\caption[Plots showing the reliability of error estimates]
	{Plots showing the reliability of error estimates.  The 
	vertical axis is the difference of two calculations relative to 
	the more precise of the two, and the horizontal axis is the 
	relative error reported by the numerical integration.  The 
	\code{Cuhre} algorithm results are shown in a) and \code{Suave} 
	results are in b).  Points with differences greater than twice 
	the reported errors are red, those with differences between one 
	and two times the error are green, and those with error estimates 
	large enough to cover their difference from the ``best'' value 
	are blue.  \comment{Points corresponding to the same integral are 
	connected with lines.  The lines are red when an increase in 
	sample size results in an increase in the difference and blue 
	when the difference decreases.  Points representing integrals 
	with an equal number of samples are coloured black.}}\label{fig.rtc2.reserr}
\end{figure}

We now turn our attention to the \code{Cuhre} algorithm, so we will 
only be considering integrals over two or more dimensions.  
The first question we will ask here is whether the error bars 
reported by the integration algorithms are sufficient.  Because we do 
not know the correct results for most of the integrals, we evaluate 
each integral with at least three different choices for the sample 
size, $N$.  Making the assumption that the calculation with the 
largest $N$ is ``correct'', we can compare the difference between each 
computed integral and the most accurate one to the error estimate 
reported by that integral.  This is shown in figure 
\ref{fig.rtc2.reserr-C}, where blue points have sufficient error bars 
and green points are within twice the error bars.  The few red 
points are the outliers which differ from their largest $N$ partners 
by more than twice their error estimates.  In a moment we will 
compare every computed integral with other calculations of the same 
integral, so the red points here will also have greater than 
$2\sigma$ difference in that comparison.  Many of these points come 
from the same integrals, so there are actually not very many 
integrals which will need to be examined in detail.  
Our choice to prefer the 
\code{Cuhre} algorithm over the adaptive Monte Carlo algorithm \code{Suave} is 
justified by figure \ref{fig.rtc2.reserr-S}, where we see that the 
\code{Suave} algorithm is as likely as not to underestimate the 
error.  In its defence, \code{Suave} frequently reports 
a 100\% $\chi^{2}$ estimate that the reported error is insufficient, 
but this is not particularly helpful in finding accurate values.

\begin{figure}
	\subfloat[]{\label{fig.rtc.err-F200}\includegraphics[width=0.45\textwidth]{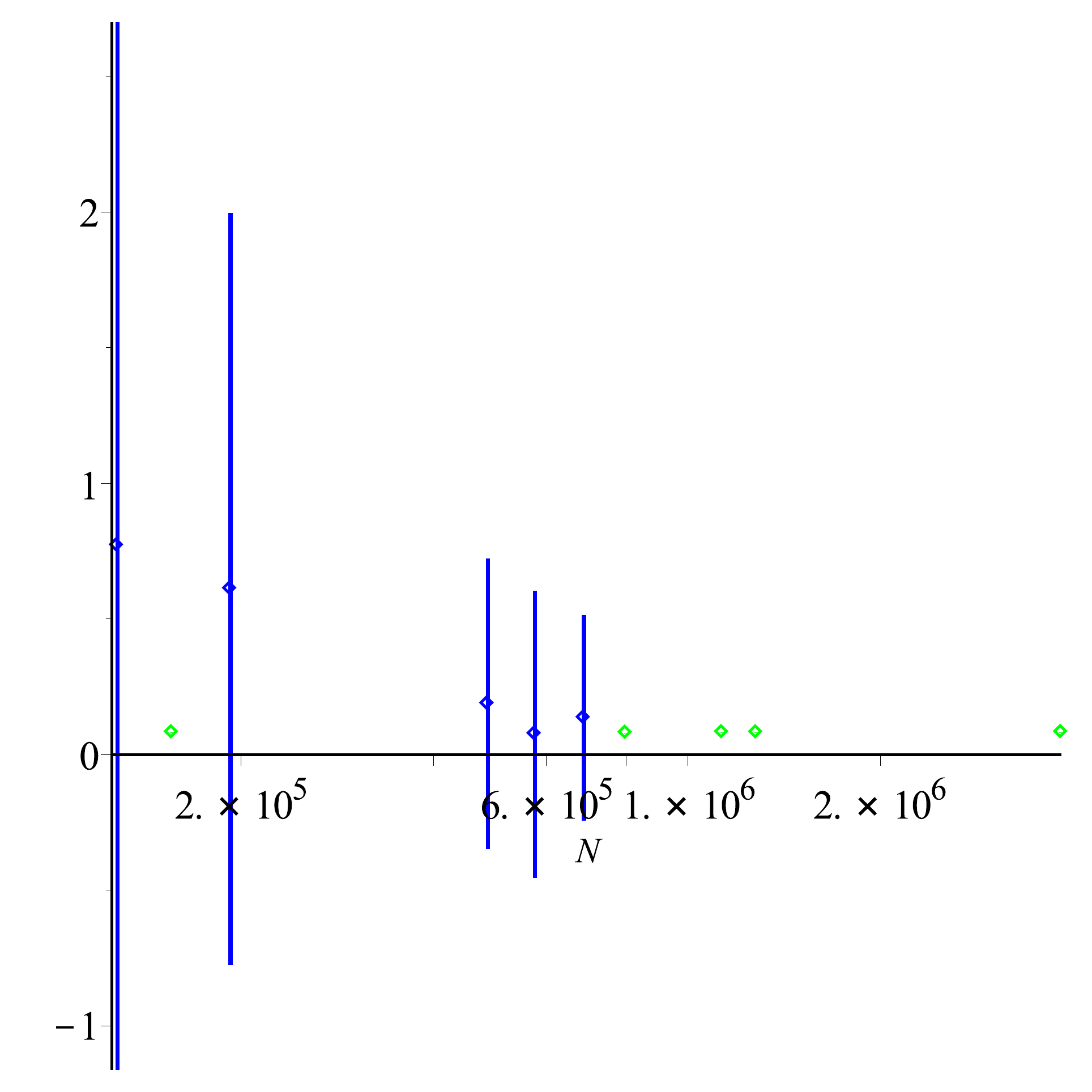}}
	\hfill
	\subfloat[]{\label{fig.rtc.err-F242}\includegraphics[width=0.45\textwidth]{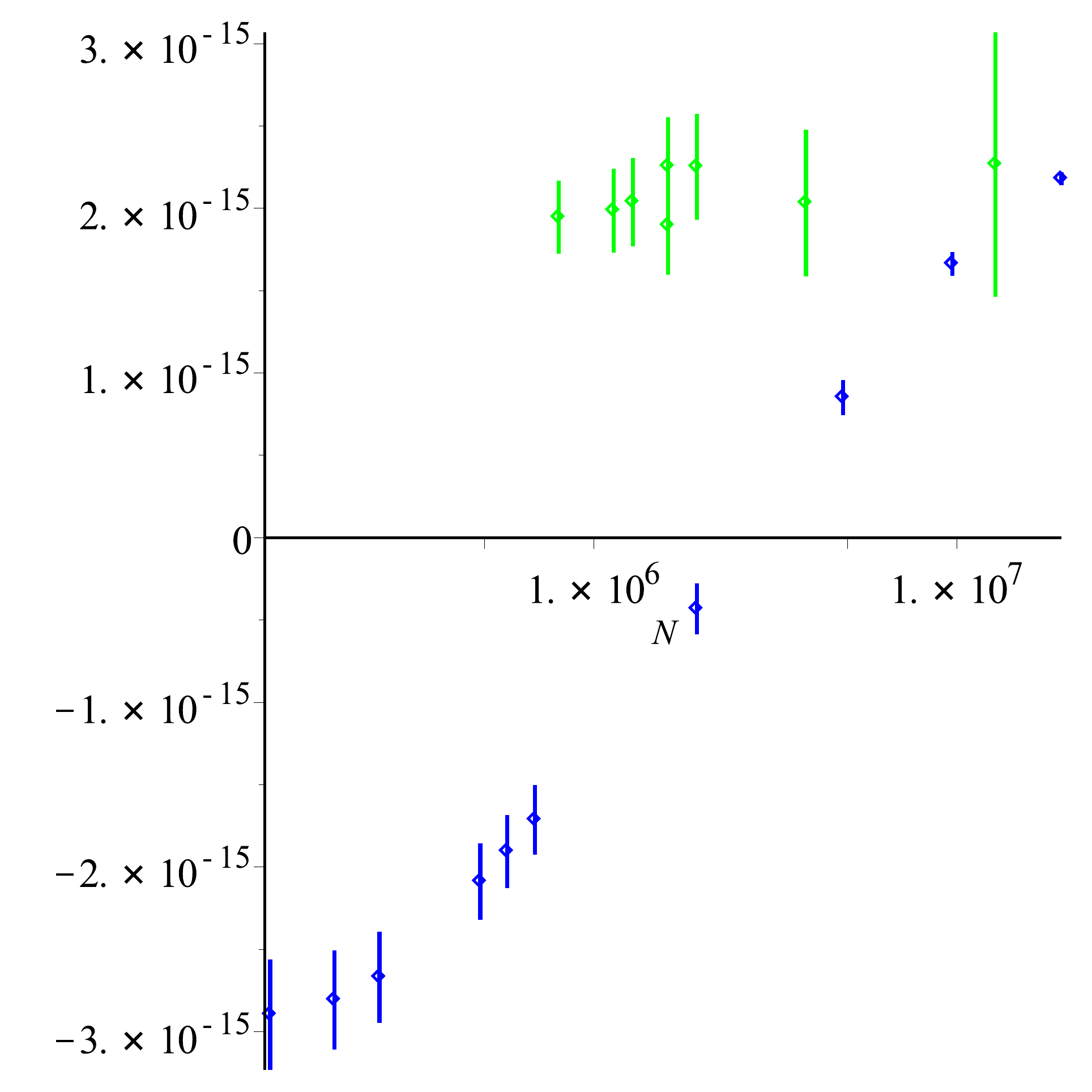}}\\
	\subfloat[]{\label{fig.rtc.err-F199}\includegraphics[width=0.3\textwidth]{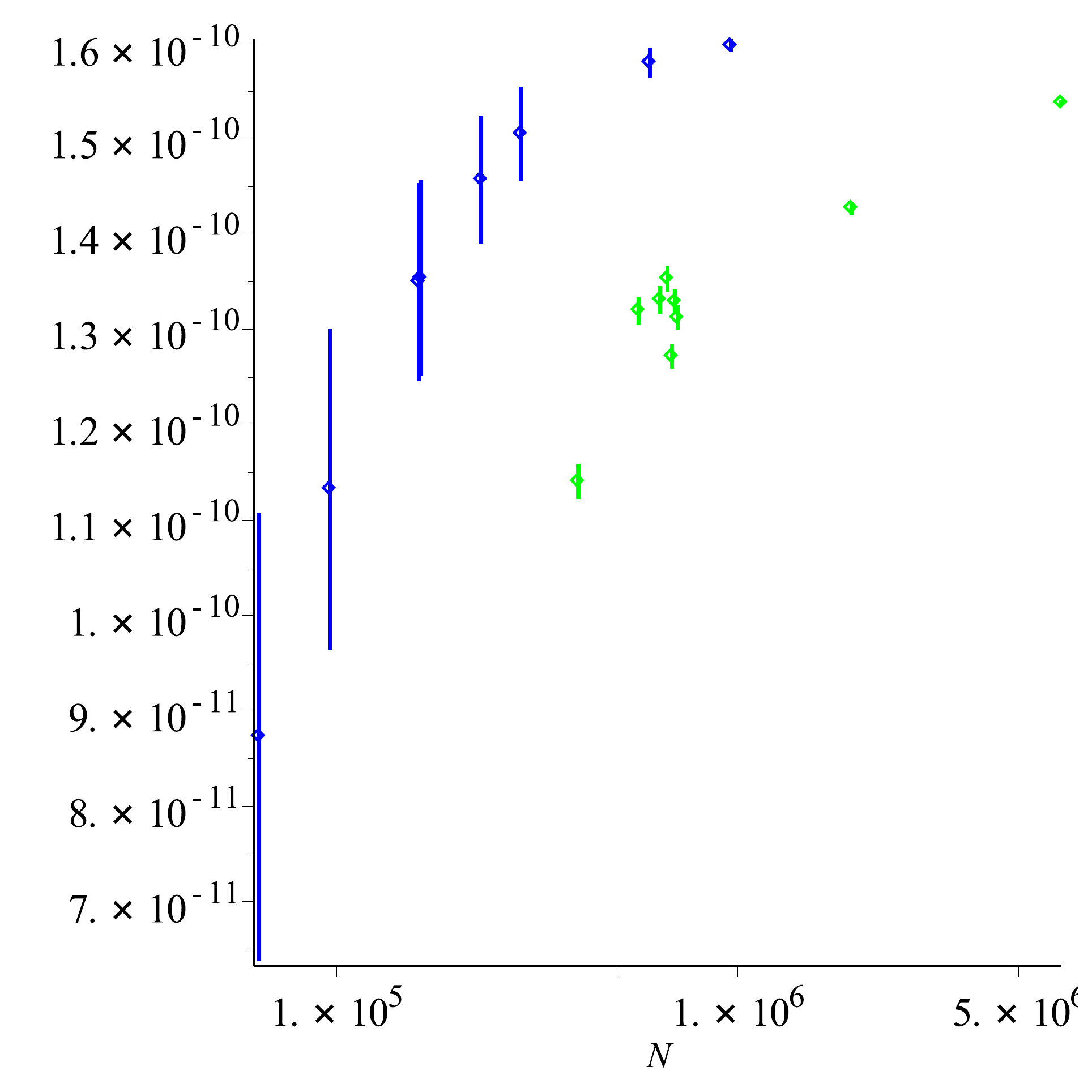}}
	\hfill
	\subfloat[]{\label{fig.rtc.err-F236}\includegraphics[width=0.3\textwidth]{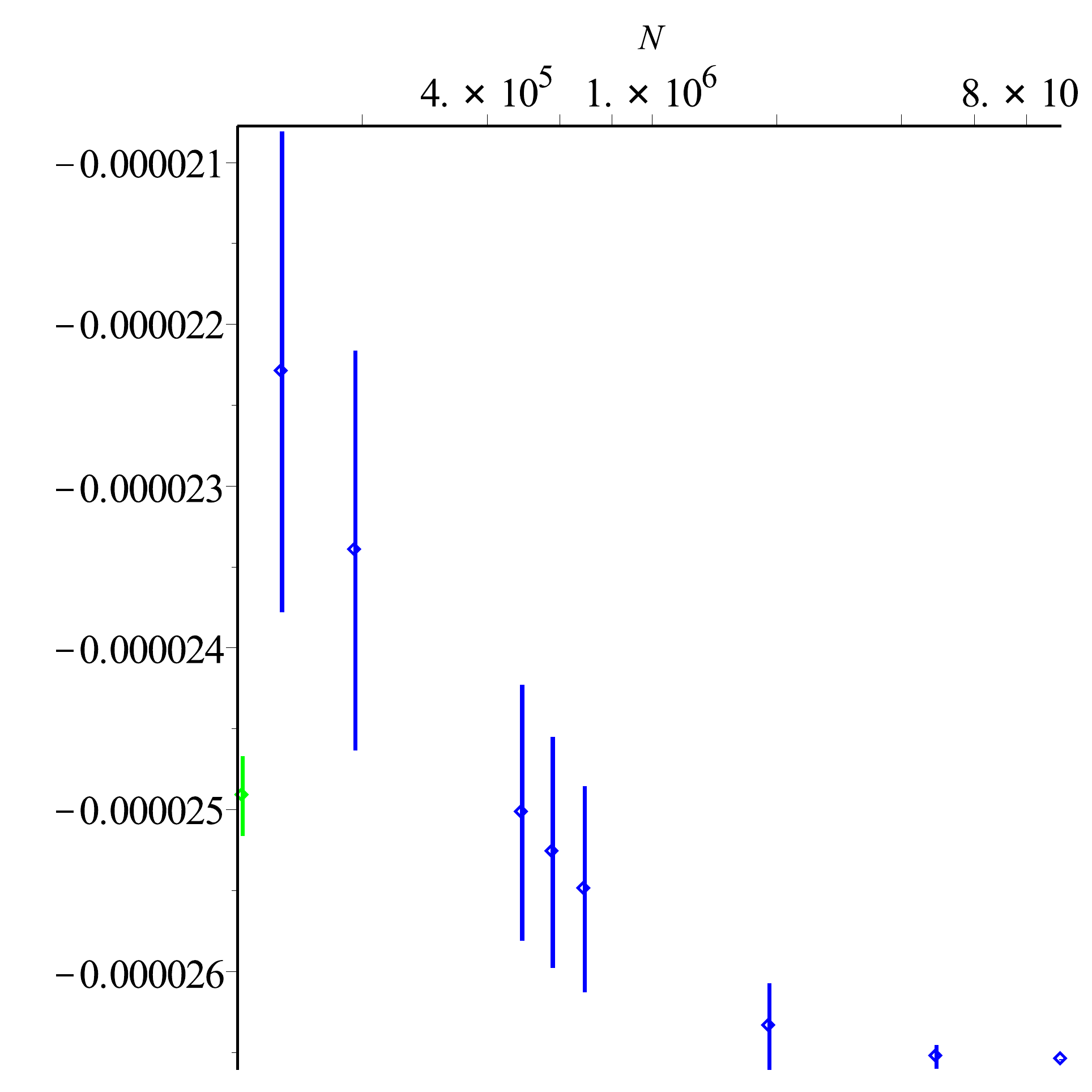}}
	\hfill
	\subfloat[]{\label{fig.rtc.err-F238}\includegraphics[width=0.3\textwidth]{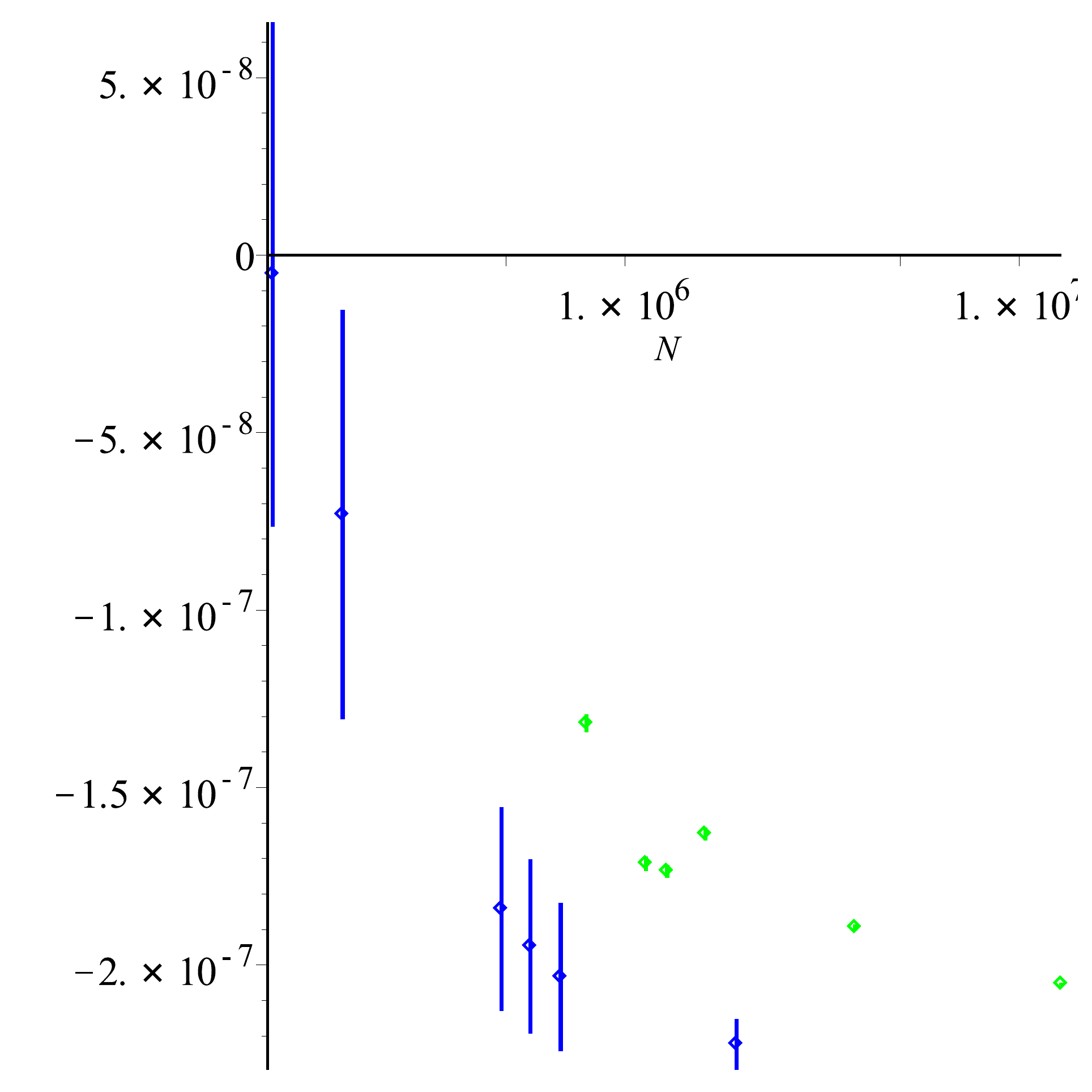}}
	\caption[Several integrals from the tachyon profile calculated 
	with different values of $N$]
	{Several integrals from the tachyon profile calculated with 
	different values of $N$.  The error bars are those reported by 
	the \code{Cuhre} algorithm.  When the \code{Suave} algorithm 
	gives results which fit on the same scale they are included as 
	the green data.}\label{fig.rtc.err-FT}
\end{figure}

The \code{Suave} algorithm can still be useful for comparison, 
however.  Its error bars are not helpful, but if the same quantity 
computed in \code{Cuhre} and \code{Suave} differs by more than a few 
percent it is worth closer examination.  This is how the two terms 
marked with asterisks in table \ref{tab.rtc.solutionmodesRC} were 
identified.  The integrals responsible for the troublesome behaviour 
of these two quantities are plotted in figures 
\ref{fig.rtc.err-F200} and \ref{fig.rtc.err-F242}.  There were some 
other terms which were flagged by this test, but they converged 
reasonably well once the sample size was increased sufficiently.

\begin{figure}
	\subfloat[]{\label{fig.rtc.err-F59}\includegraphics[width=0.3\textwidth]{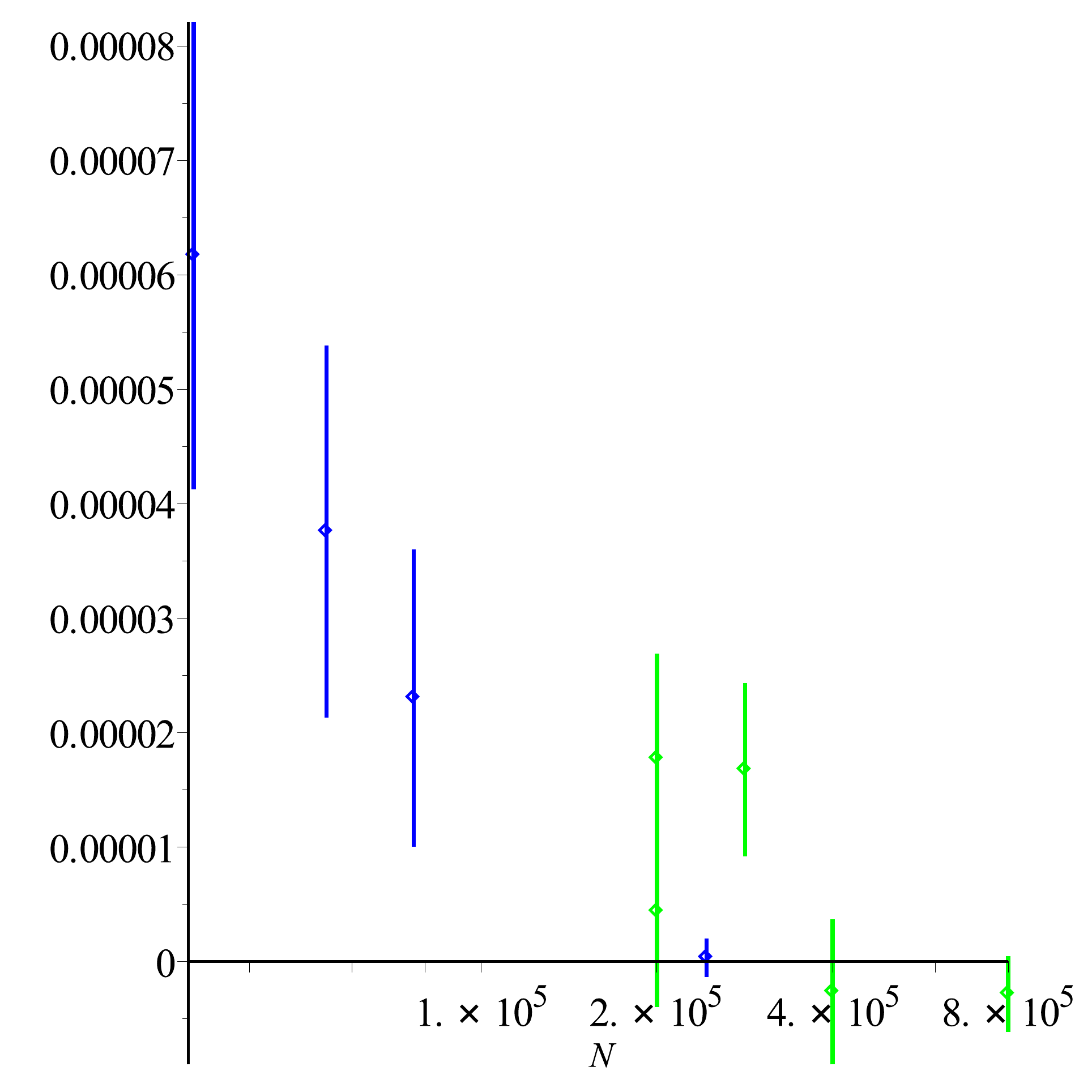}}
	\hfill
	\subfloat[]{\label{fig.rtc.err-F92}\includegraphics[width=0.3\textwidth]{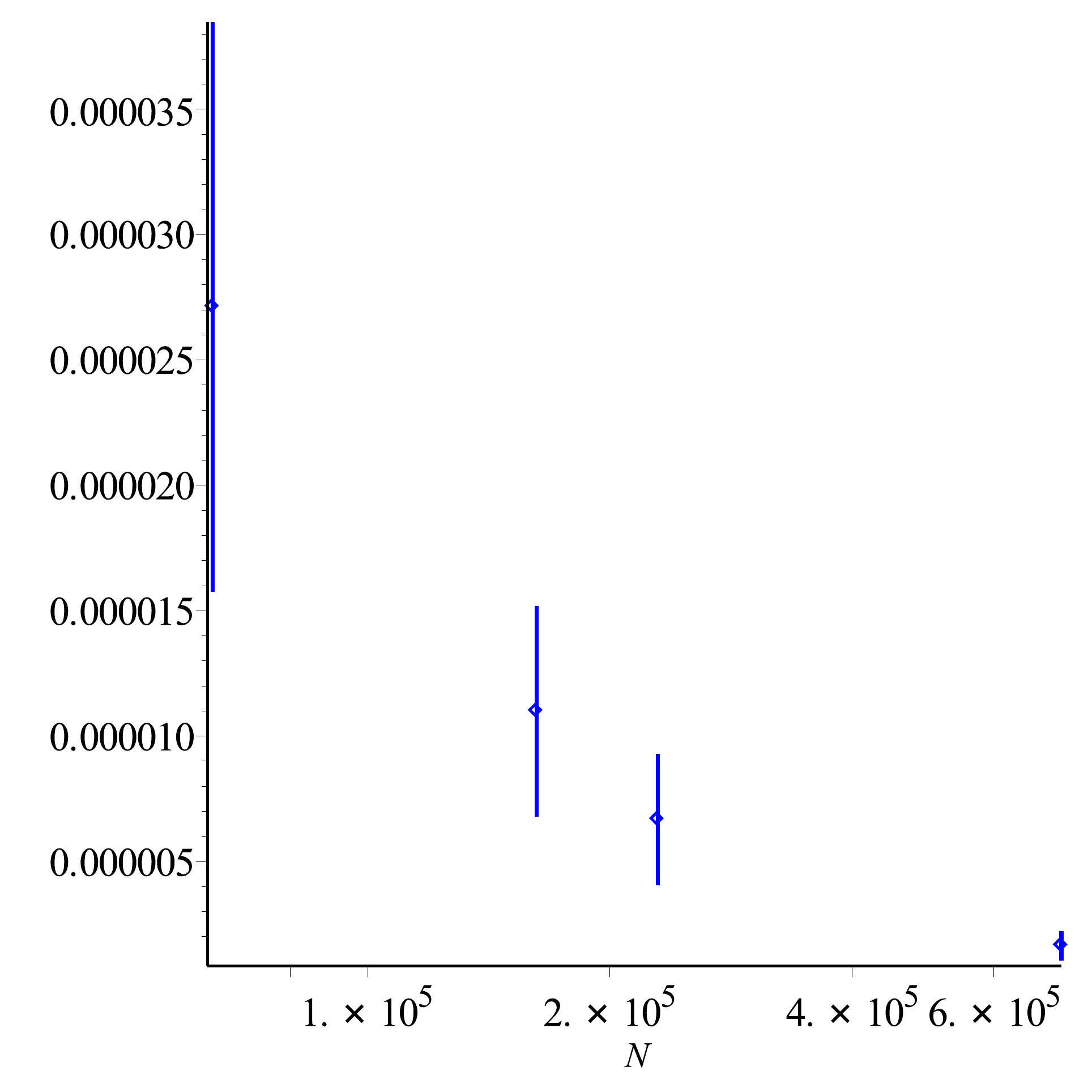}}
	\hfill
	\subfloat[]{\label{fig.rtc.err-F97}\includegraphics[width=0.3\textwidth]{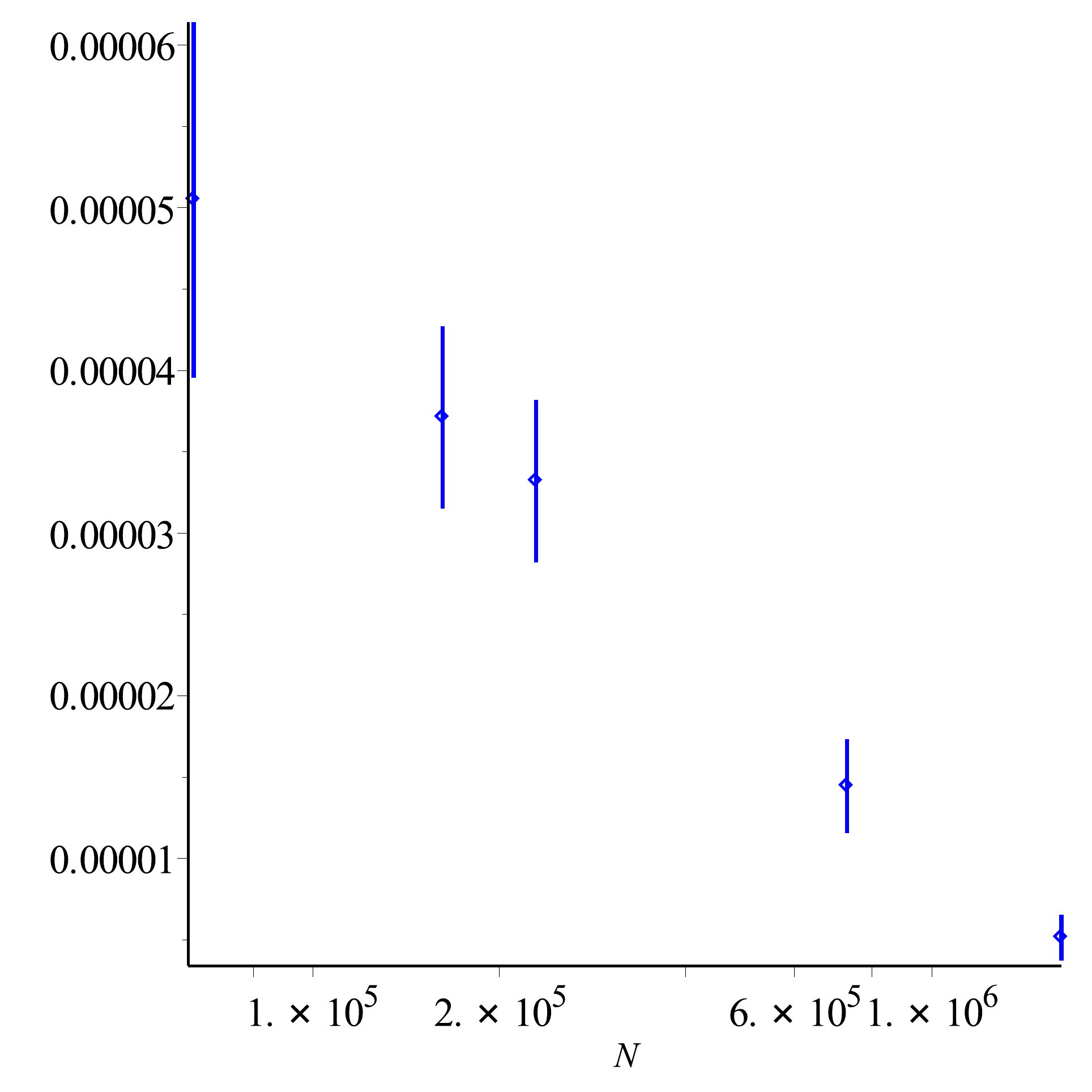}}\\
	\subfloat[]{\label{fig.rtc.err-F98}\includegraphics[width=0.3\textwidth]{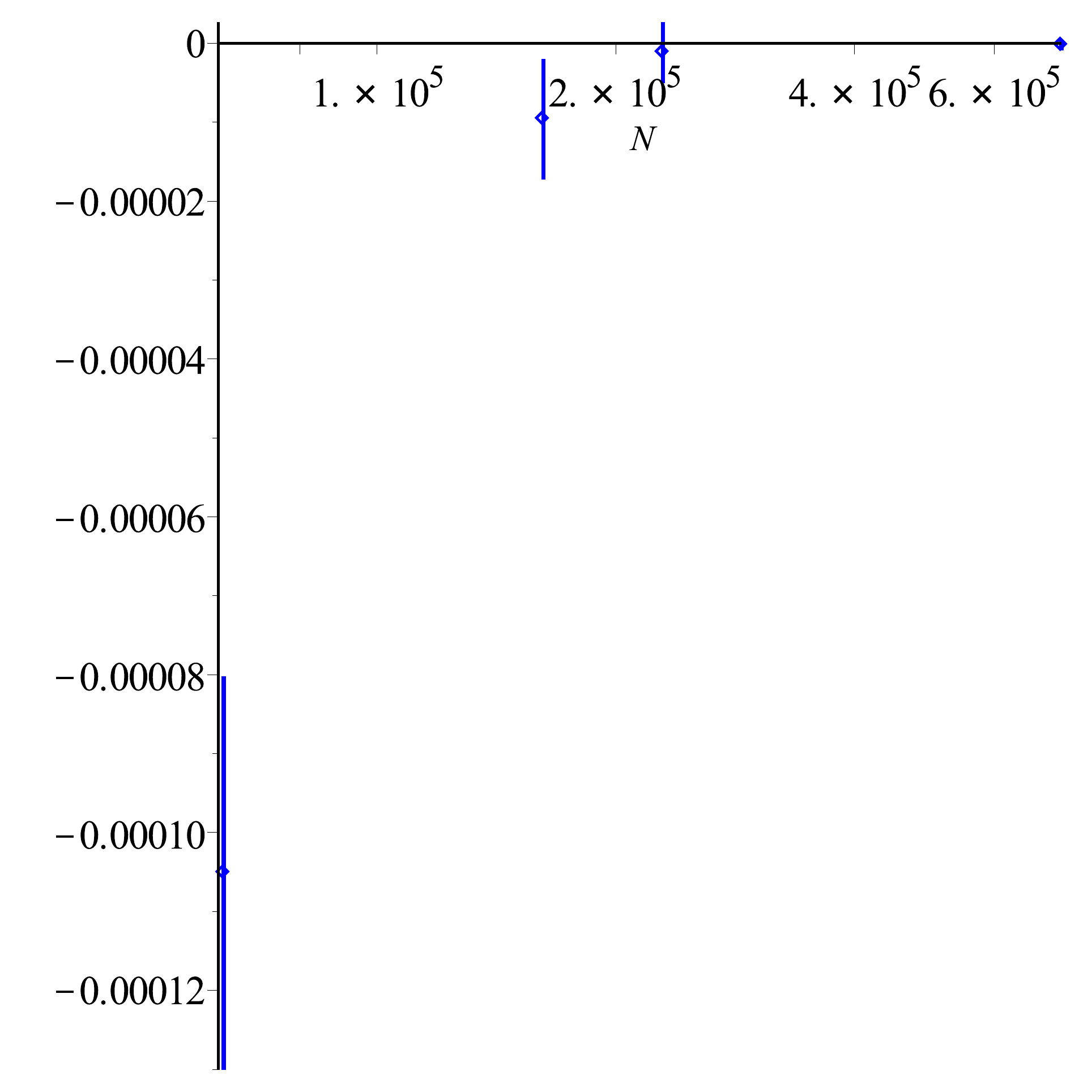}}
	\hfill
	\subfloat[]{\label{fig.rtc.err-F99}\includegraphics[width=0.3\textwidth]{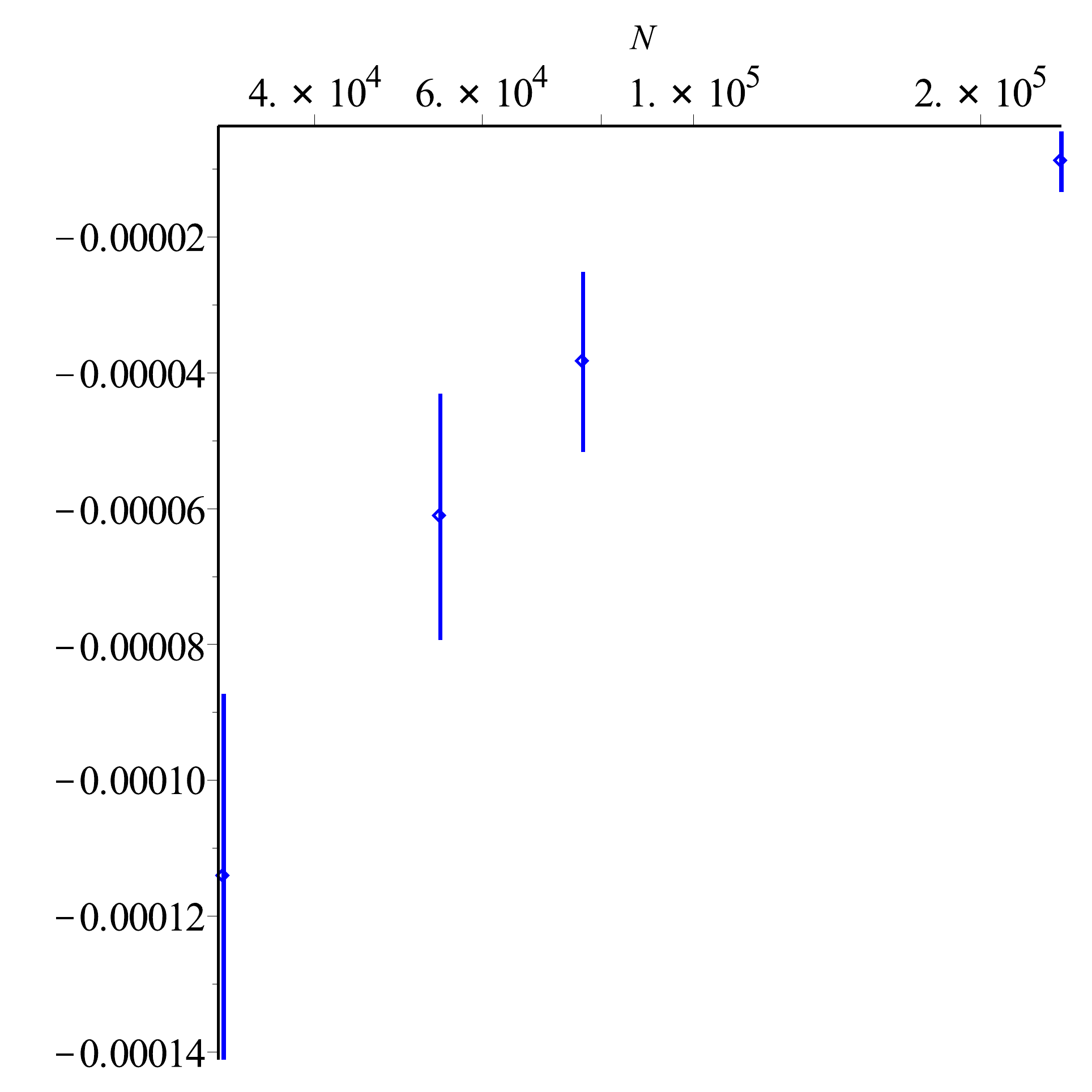}}
	\hfill
	\subfloat[]{\label{fig.rtc.err-F101}\includegraphics[width=0.3\textwidth]{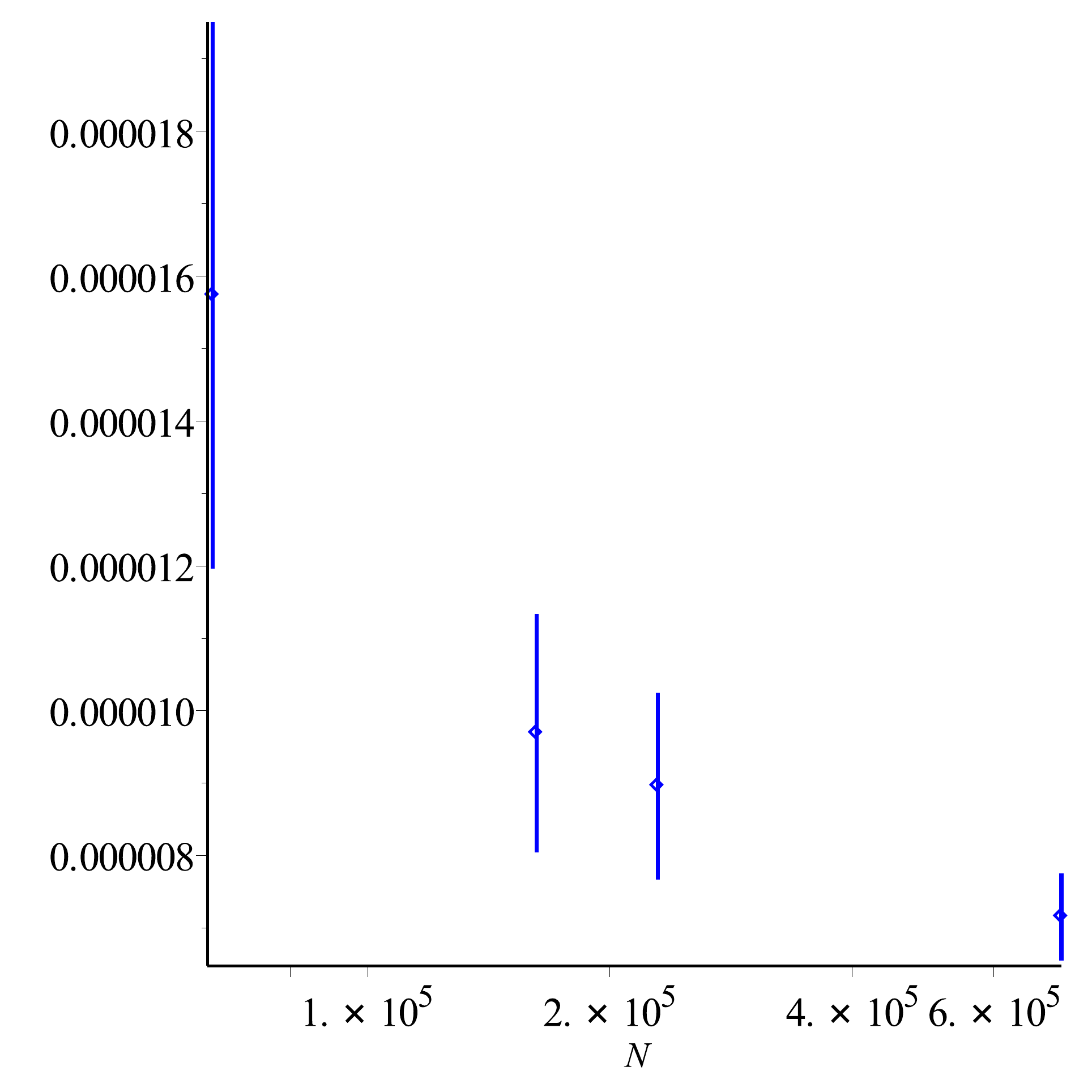}}\\
	\subfloat[]{\label{fig.rtc.err-F116}\includegraphics[width=0.3\textwidth]{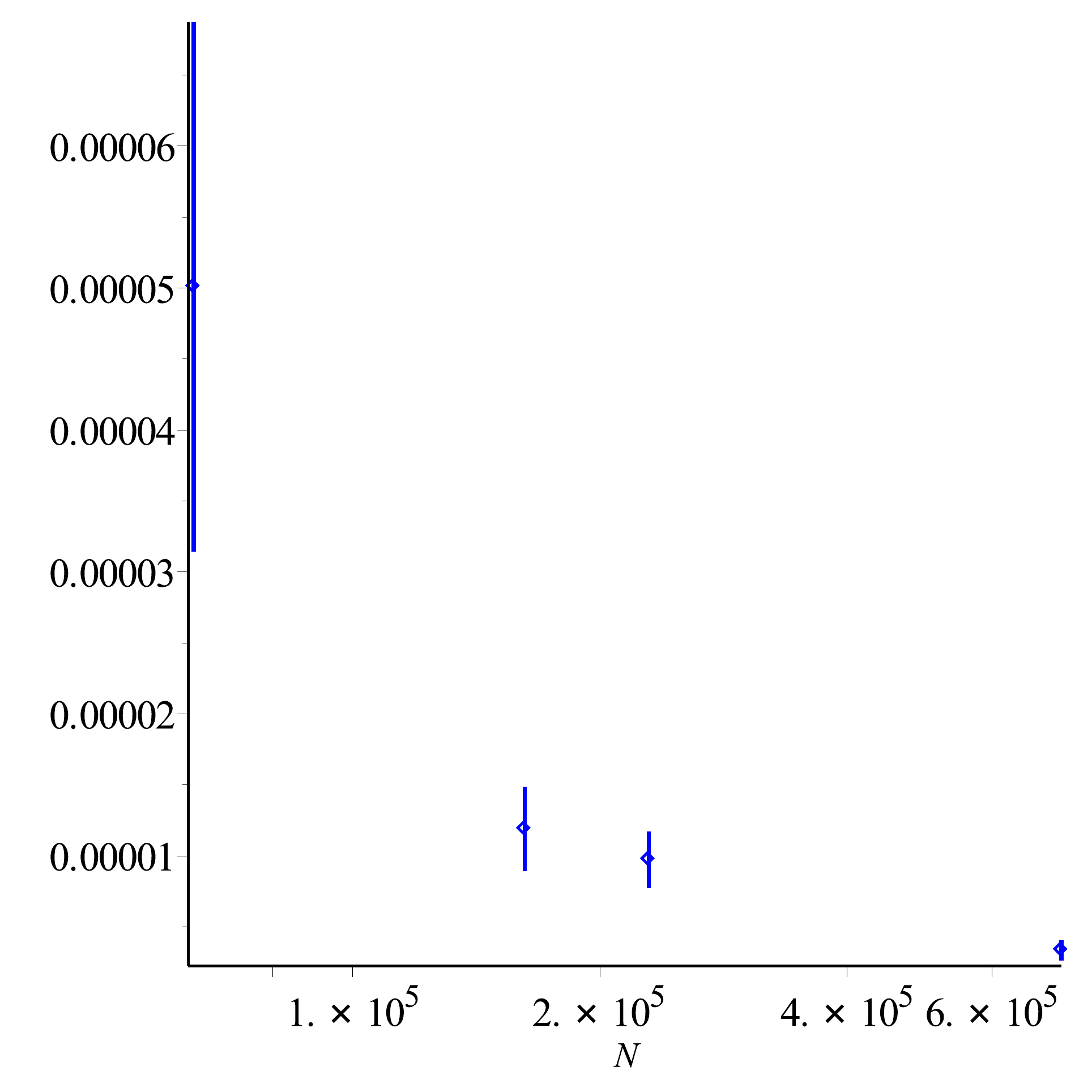}}
	\hfill
	\caption[Several integrals from the consistency checks calculated 
	with different value of $N$]
	{Several integrals from the consistency checks calculated 
	with different value of $N$.  The error bars are those reported by 
	the \code{Cuhre} algorithm.  When the \code{Suave} algorithm 
	gives results which fit on the same scale they are included as 
	the green data.}\label{fig.rtc.err-FC}
\end{figure}

With many integrals each calculated for several different values of 
$N$, we have a sizeable collection of data to examine.  Among the 848 
\code{Cuhre} integrals, there is only one instance of the error 
estimate increasing as the sample size was increased, so we can 
safely say that the reported error bars decrease monotonically as 
$N\rightarrow\infty$.  We can then examine the quantity 
$\frac{|x_{i}-x_{j}|}{\sqrt{\Delta x_{i}^{2}+\Delta x_{j}^{2}}}$ for 
every pair of calculations of the same integral.  We find that 88\% 
of pairs are within $\sigma$ of each other, while 94\% are within 
$2\sigma$.  Those which disagree by more than $2\sigma$ can be 
studied individually, since they correspond to only 15 different 
integrals.  Of those, four only disagree due to a single computation 
each with very low $N$ (about 500 samples) that has a 50\% $\chi^{2}$ 
chance of being incorrect.  One of the remaining 11 integrals is also 
the one responsible for the tachyon profile coefficient 
$\beta_{6}^{(0)}$, so adding in the integral responsible for the 
other flagged term in table \ref{tab.rtc.solutionmodesRC} we have 12 
integrals to examine.  These are plotted with various values of $N$ 
in figures \ref{fig.rtc.err-FT} and \ref{fig.rtc.err-FC}.  The values 
and their error bars are shown in blue, and when appropriate to the 
scale of the plot the corresponding \code{Suave} results are also 
shown in green. 

The first two plots represent the parts of the tachyon profile for 
which we used the \code{Suave} results.  In figure \ref{fig.rtc.err-F200} 
the problem is not that the results are inconsistent, but that the 
errors are so large that the results are meaningless.  It looks 
likely that as $N$ is increased the results will continue to converge 
to something quite close to the \code{Suave} result.  For the other 
integral, figure \ref{fig.rtc.err-F242} shows that as $N$ is made 
extremely large we finally find something like the much more 
consistent \code{Suave} results.  The slope, however, does not yet 
appear to be significantly slowing down, so we cannot be sure that it is 
convergent.  The rest of the plots in figures 
\ref{fig.rtc.err-F199}-\ref{fig.rtc.err-F238} show the other 
integrals that contribute to the tachyon profile and have more than 
a $2\sigma$ variation between points.  They all show signs that once 
$N$ is sufficiently large they converge quite well.  Only for smaller 
$N$ do the error estimates appear to be insufficient.  

Moving on to integrals which contribute to consistency checks, in 
figure \ref{fig.rtc.err-FC} we see that the majority are fine.  
Only \ref{fig.rtc.err-F97} does not appear to converge.  As with the 
examples shown in figure \ref{fig.rtc.err-FT}, this may well be 
linear until some critical $N$ where it begins to converge.  In 
addition, while only the quantities composed of small sums of these 
integrals are supposed to vanish, in many cases each of the integrals 
will vanish independently.  These seven examples all become closer to 
vanishing as $N$ is increased, and only \ref{fig.rtc.err-F101} is not 
getting very close to zero.  Convergence of these results does not 
appear to be very much of an issue, and I expect that if we continued 
to increase $N$ by another factor of ten they would all continue to 
approach zero.

\subsection{Roundoff errors}

Each integrand may contain a number of terms which are divergent 
either on the boundary of the region, $t_{i}\in\{a,b\}$, or on a 
diagonal, $t_{i}=t_{j}$.  While the renormalization is designed 
specifically so 
that these divergences will cancel, individual terms evaluated near 
these regions can be very large.  Because we are limited to the double 
precision floating point datatype, each term has a relative precision 
of approximately $10^{-16}$.  Any time an individual term is more 
than $10^{16}$ times the theoretical value of the integrand evaluated 
at the same point, the machine uncertainty coming from that term can 
dominate the result.  We would hope that since this only happens for 
a small subset of the points sampled the effect will be negligible 
as the number of points increases, but this is not the case.  If we 
take a random sample of $N$ points, as is done for Monte Carlo 
integration, we would expect the closest point 
to a given boundary (or other codimension 1 subspace) to be 
$\sim\frac{1}{N}$ away.  Individual terms, however, often have a 
$\frac{1}{t^{2}}$ divergence from the OPE of the marginal operator, 
which would lead to 
$\sim N^{2}$ divergence for the closest point.  This grows faster 
than the denominator, $N$, so the roundoff error in the resulting 
integral should increase linearly with the number of points.  The 
deterministic case is actually worse because some points are 
intentionally chosen near or even right on the boundary.  To combat 
this, whenever a sample point is close to a boundary or a diagonal, 
we can replace it with a nearby point giving a decent approximation 
to the integrand.  The integrand function is effectively replaced by 
one where the value is held constant on small strips.  While this 
means that a perfect integration with no 
uncertainty would give an incorrect result, the errors introduced 
this way are less problematic than the roundoff errors when we sample 
many points without any regulation.    

When we discussed the differences between the big $G$ and little $g$ 
renormalization schemes, we saw that the lack of a regulator was 
an advantage of the little $g$ scheme.  Here we have introduced 
another regulator, so we naturally ask why this is not a problem.  
The regulator in the big $G$ scheme was required by the theory in 
that scheme, and we wanted the limit as it approached zero.  This 
regulator is to prevent roundoff error, 
which is the unavoidable result of using a floating point datatype.  
Since we are regulating the integration region anyway, we might ask 
why (aside from issues regarding finiteness at higher orders) we did 
not use the big $G$ scheme.  By using the little $g$ scheme, the 
integrand is independent of the regulator, which simplifies the 
integration process.  There is not a different integrand for each 
value of the regulator, and instead of a limit, we only use a small 
value of the regulator, namely $3\E{-4}$, for which the integrands 
always evaluate with negligible errors.  The choice of regulator for 
these roundoff errors is arbitrary, but we can estimate the error we 
have introduced by using the same regulator to replace the value of 
the integrand with zero near boundaries and diagonals.  Fortunately, 
the differences are minor compared with the statistical errors which 
are accounted for by the algorithms' reported uncertainties.

\subsection{Consistency checks} \label{sec.rtc.checks}

Since the programs to construct wedge states with insertions and 
produce and evaluate integrals corresponding to the tachyon profile 
are quite complicated, it is worth using them to evaluate some known 
quantities.  We will see that the numerical integration process 
gives results which are consistent with expectations the majority of 
the time, despite the presence of counterterms and the uncertain 
nature of numerical integration.  
An obvious choice for a quantity which we know 
is the equation of motion, which should vanish.\comment{ for 
any form of the solution \eq{eq.rtc.solutions}.}  The equation of 
motion, however, has ghost number two, which means that its 
expectation value by itself will trivially vanish because the ghosts 
are not saturated.  In order to test that $Q_{B}\Psi+\Psi*\Psi$ 
vanishes we test that its overlaps with various other 
string fields all vanish.  Because the equation of motion is stronger 
than just requiring that the equation of motion annihilates all 
states and actually tells us that it should vanish exactly, as long 
as the string field is constructed properly these correlation 
functions should work out to zero whether they themselves were 
computed correctly or not.  In order to test that a non-trivial 
result also gives the correct answer, we look to the action.  Because 
this is an exactly marginal solution, we expect the energy to 
vanish, and because the energy is proportional to the action, the 
action should vanish as well.  The action has ghost number three and 
does not need any additional test states inserted.  That this also 
vanishes is our first strong test that non-trivial expectation values 
are computed successfully.  A summary of these test calculations and 
their results using the deterministic algorithms is found in tables 
\ref{tab.rtc.checks1} and \ref{tab.rtc.checks2} at the end of this 
section, and all of them are 
expected to vanish.  The majority of the 
results are consistent with zero, but a few exceptions require 
detailed examination.  These six examples are in table 
\ref{tab.rtc.checks3}, which restates their values using the 
deterministic algorithms and then includes the corresponding results 
with Monte Carlo calculations and with deterministic calculations 
using a vanishing integrand near borders and diagonals where 
cancelling singularities may occur.  

The values in tables \ref{tab.rtc.checks1} and \ref{tab.rtc.checks2} all 
use a border with width $\epsilon=3\cdot10^{-4}$.  When the integrand is 
sampled within $\epsilon$ of a boundary or a diagonal, the closest 
point on the edge of this strip is used instead.  In the last column 
of table \ref{tab.rtc.checks3}, when the integrand was sampled at 
points within these strips, zero was returned instead.  The 
difference between these results gives an estimate of how important 
the regulated region is to the final result of the integral, and we 
can see that in most cases it is small compared to the error estimates. 

\begin{table}[tb]\centering\begin{tabular}{|>{$}c<{$}|>{$}c<{$}|>{$}c<{$}|>{$}c<{$}|}
\hline
\text{Quantity}&\text{\code{Cuhre}/\code{QAG}}&\text{\code{Suave}}&\text{\code{Cuhre}/\code{QAG} with 0}\\
\hline
\langle ce^{X^{(0)}},\text{EOM}^{(3)}\rangle&(-6.1\pm0.6)\E{-11}&(4.1\pm0.4)\E{-4}&(-6.1\pm0.6)\E{-11}\\
\langle\Psi^{(1)},\text{EOM}^{(3)}\rangle&(-8.5\pm0.8)\E{-11}&(6.1\pm0.7)\E{-4}&(-8.5\pm0.8)\E{-11}\\
\langle\Psi,Q_{B}\Psi\rangle^{(4)}&(-8.7\pm1.2)\E{-10}&(-1.5\pm1.3)\E{-2}&(-8.7\pm1.2)\E{-10}\\
\frac{\partial}{\partial C^{L}}\langle ce^{3X^{0}},\text{EOM}^{(5)}\rangle&(6.2\pm2.1)\E{-5}&(-0.3\pm1.1)\E{-6}&(-0.4\pm1.3)\E{-4}\\
\frac{\partial}{\partial C_{0}}\langle c,\text{EOM}^{(6)}\rangle&(-2.2\pm0.5)\E{-4}&(8.7\pm7.8)\E{-4}&(0.1\pm2.1)\E{-2}\\
\frac{\partial}{\partial C_{1}}\langle\Psi^{(3)},\text{EOM}^{(3)}\rangle&(-3.0\pm0.2)\E{-8}&(-5.1\pm0.7)\E{-4}&(0.8\pm1.7)\E{-3}\\
\hline
\end{tabular}
\caption[Numerical results for consistency checks which require 
further analysis]
{Numerical results for consistency checks which require 
further analysis.  Among the results which are expected to vanish, 
these six have deterministic results which do not.  They are given 
using the standard deterministic algorithms \code{Cuhre} and 
\code{QAG}, using the adaptive 
Monte Carlo algorithm \code{Suave}, and using the deterministic 
algorithms with the integrand replaced with 0 on the regulated strips 
near potential singularities instead of using the value of the 
integrand at a safe nearby point.}
\label{tab.rtc.checks3}\end{table}

\begin{table}[t]\centering\begin{tabular}{|c|c|>{$}c<{$}|c|c|c|c|}
	\hline
	Algorithm & \text{Quantity} & \text{Result} & \multicolumn{4}{c|}{Dimensions and Sample Sizes}\\
	\hline
	\multirow{4}{*}{\code{Cuhre}}&
	\multirow{4}{*}{$\frac{\partial}{\partial C^{L}}\langle ce^{3X^{0}},\text{EOM}^{(5)}\rangle$}
	&(6.2\pm2.1)\E{-5}&\multirow{4}{*}{2}&16055&\multirow{4}{*}{3}&32131\\
	&&(3.7\pm1.6)\E{-5}&&18005&&54229\\
	&&(2.3\pm1.3)\E{-5}&&25545&&76581\\
	&&(0.2\pm1.7)\E{-6}&&81055&&243205\\
	\hline
	\multirow{4}{*}{\code{Cuhre}}&
	\multirow{4}{*}{$\frac{\partial}{\partial C_{0}}\langle c,\text{EOM}^{(6)}\rangle$}
	&(-2.2\pm0.5)\E{-4}&\multirow{4}{*}{3}&32131&\multirow{4}{*}{4}&64107\\
	&&(-7.1\pm2.6)\E{-5}&&54229&&162027\\
	&&(-4.0\pm1.7)\E{-5}&&76581&&229347\\
	&&(-9.1\pm5.2)\E{-6}&&243205&&729045\\
	\hline
	\multirow{4}{*}{\code{Cuhre}}&
	\multirow{4}{*}{$\frac{\partial}{\partial C_{1}}\langle\Psi^{(3)},\text{EOM}^{(3)}\rangle$}
	&(-2.96\pm0.19)\E{-8}&\multirow{4}{*}{3}&32131&\multicolumn{2}{c}{}\\
	&&(-2.99\pm0.13)\E{-8}&&54229&\multicolumn{2}{c}{}\\
	&&(-2.99\pm0.11)\E{-8}&&76581&\multicolumn{2}{c}{}\\
	&&(-3.01\pm0.06)\E{-8}&&243205&\multicolumn{2}{c}{}\\
	\cline{1-5}
\end{tabular}
\caption[Three consistency checks shown with several different sample sizes]
{Three of the consistency checks in table \ref{tab.rtc.checks3} are 
shown with different sample sizes.  The first two are 
computed as the sum of two integrals with different dimensions, while 
the third is a single integral.  We expect to see results which tend 
towards zero as the sample size is increased.}
\label{tab.rtc2.3checks-n}\end{table}  

Looking at table \ref{tab.rtc.checks3}, the first two quantities, 
$\langle ce^{X^{(0)}},\text{EOM}^{(3)}\rangle$ and 
$\langle\Psi^{(1)},\text{EOM}^{(3)}\rangle$, have very similar 
behaviours because the second integrand is $\sqrt{2}$ times the 
first.  They are one dimensional 
integrals, so we can do them analytically and find that the results 
are exactly zero.  The integrands for these two are 
increasing as they approach each of the boundaries, which would 
suggest that the discrepancy comes from the regulated region, but 
the results with 0 inserted near the boundaries are identical.
In fact, the \code{QAG} algorithm always seems to give identical 
results with either choice of regulation near the boundaries, 
suggesting that that it does not pick points too close to the limits 
of integration.  
If we redo these integrals requesting much higher accuracy, however, 
we find results which are less precise but consistent with zero.  
Perhaps requesting a higher accuracy causes the algorithm to notice 
the cusp where the regulated strips at the boundary are, and that 
increases the error estimate.  
The kinetic energy at fourth order is a peculiar case of rounding 
errors, and it was already discussed in the context of convergence.  
None of the integration 
algorithms is correct all of the time, and we should not expect them 
to be, but \code{QAG} is problematic because it gives less control 
over the sample size and does not report the total number of points 
it uses.  We cannot show these results with different sample sizes, 
as we do for other quantities of interest.  Because these are 
one-dimensional integrals, however, we can expect that the majority 
of them will be computed very accurately.

For the other three quantities, the integrals are multidimensional 
so we can evaluate them using the \code{Cuhre} algorithm with several 
different sample sizes and see if they become closer to zero.  This 
is shown in table \ref{tab.rtc2.3checks-n}.  The first two of these 
both have error estimates that decrease as the sample size is 
increased, and values that decrease even faster.  For the first, we 
see agreement once the sample size is large enough, and for the 
second we can suspect that the result will continue to tend towards 
zero.  The most important integrals in these two quantities were 
shown in figures \ref{fig.rtc.err-F59} and \ref{fig.rtc.err-F98} 
respectively, where we can see the convergence to 0 as $N$ is 
increased.  
In the case of the final quantity, however, the result clearly 
does not vanish.  If we change the thickness of the regulating 
border, however, it causes a significant fluctuation, and for 
extremely thin borders both the result and the uncertainty become 
much larger.  This suggests that it is a genuine case of the border 
region having a significant impact.  Unfortunately, the only way to 
resolve this issue would be to use higher precision floating point 
datatypes.

\afterpage{\clearpage}

\begin{table}\small\begin{tabular}{|>{$}c<{$}|>{$}p{.78\textwidth}<{$}|}
\hline
\langle c,\text{EOM}^{(2)}\rangle&0\\
\langle ce^{2X^{0}},\text{EOM}^{(2)}\rangle&0\\
\langle ce^{X^{0}},\text{EOM}^{(3)}\rangle&-(6.1\pm0.6)\E{-11}\\
\langle ce^{3X^{0}},\text{EOM}^{(3)}\rangle&0\pm1.1\E{-16}\\
\langle c,\text{EOM}^{(4)}\rangle&-(0.6\pm1.9)\E{-6}-(0.8\pm1.3)\E{-6}C^{L}\\
\langle ce^{2X^{0}},\text{EOM}^{(4)}\rangle&(1.7\pm2.5)\E{-6}+(9.2\pm9.7)\E{-7}C^{L}\\
\langle ce^{4X^{0}},\text{EOM}^{(4)}\rangle&(2.8\pm8.7)\E{-11}\\
\langle ce^{X^{0}},\text{EOM}^{(5)}\rangle&(1.8\pm2.5)\E{-5}-(1.3\pm1.8)\E{-5}C^{L}-(0.2\pm3.2)\E{-4}(C^{L})^{2} + (2.1\pm2.0)\E{-5}C_{1}+(0.6\pm2.3)\E{-5}C_{0}\\
\langle ce^{3X^{0}},\text{EOM}^{(5)}\rangle&(0.2\pm2.3)\E{-5}+(6.2\pm2.1)\E{-5}C^{L}-(0.3\pm2.7)\E{-5}C_{1}-(0.7\pm6.6)\E{-6}C_{0}\\
\langle ce^{5X^{0}},\text{EOM}^{(5)}\rangle&(0.2\pm3.8)\E{-11}\\
\langle c,\text{EOM}^{(6)}\rangle&-(0.8\pm2.0)\E{-3}+(0.7\pm1.9)\E{-4}C^{L}-(1.6\pm5.4)\E{-5}(C^{L})^{2}-(1.8\pm2.3)\E{-4}C_{1}+(0.2\pm7.1)\E{-3}C^{L}C_{1}-(2.2\pm0.5)\E{-4}C_{0}+(1.9\pm6.5)\E{-5}C_{0}C^{L}\\
\langle ce^{2X^{0}},\text{EOM}^{(6)}\rangle&-(0.1\pm1.3)\E{-3}+(0.7\pm3.7)\E{-4}C^{L}+(0.2\pm1.9)\E{-3}(C^{L})^{2}+(0.1\pm2.1)\E{-3}C_{1}+(0\pm1.5\E{-2})C^{L}C_{1}+(0.9\pm5.1)\E{-4}C_{0}-(0.1\pm2.9)\E{-3}C^{L}C_{0}\\
\langle ce^{4X^{0}},\text{EOM}^{(6)}\rangle&0\pm1.1\E{-5}+(1.4\pm1.0)\E{-6}C^{L}-(0.2\pm5.9)\E{-6}C_{1}+(0.4\pm8.9)\E{-7}C_{0}\\
\langle ce^{6X^{0}},\text{EOM}^{(6)}\rangle&-(0.6\pm1.3)\E{-14}\\
\langle\Psi^{(1)},\text{EOM}^{(2)}\rangle&0\\
\langle\Psi^{(2)},\text{EOM}^{(2)}\rangle&0\\
\langle\Psi^{(3)},\text{EOM}^{(2)}\rangle&0\\
\langle\Psi^{(4)},\text{EOM}^{(2)}\rangle&0\\
\langle\Psi^{(5)},\text{EOM}^{(2)}\rangle&0\\
\langle\Psi^{(1)},\text{EOM}^{(3)}\rangle&-(8.5\pm0.8)\E{-11}\\
\langle\Psi^{(2)},\text{EOM}^{(3)}\rangle&0\\
\langle\Psi^{(3)},\text{EOM}^{(3)}\rangle&-(0.5\pm1.3)\E{-5}-(0.1\pm3.4)\E{-6}C^{L}-(3.0\pm0.2)\E{-8}C_{1}\\
\langle\Psi^{(4)},\text{EOM}^{(3)}\rangle&0\\
\langle\Psi^{(1)},\text{EOM}^{(4)}\rangle&0\\
\langle\Psi^{(2)},\text{EOM}^{(4)}\rangle&-(0.3\pm2.4)\E{-3}+(2.3\pm5.2)\E{-5}C^{L}-(0.2\pm1.8)\E{-5}(C^{L})^{2}\\
\langle\Psi^{(3)},\text{EOM}^{(4)}\rangle&0\\
\langle\Psi^{(1)},\text{EOM}^{(5)}\rangle&(2.4\pm3.9)\E{-5}-(1.9\pm2.5)\E{-5}C^{L}-(0.3\pm4.5)\E{-4}(C^{L})^{2}+(3.0\pm2.8)\E{-5}C_{1}+(0.8\pm3.2)\E{-5}C_{0}\\
\langle\Psi^{(2)},\text{EOM}^{(5)}\rangle&0\\
\langle\Psi^{(1)},\text{EOM}^{(6)}\rangle&0\\
\hline
\end{tabular}
\caption[Deterministic tests of the equation of motion 
for the rolling tachyon]{Deterministic tests of the equation of 
motion for the rolling tachyon.  
Superscripts represent the order in $\lambda$ of each quantity. 
\code{Cuhre}/\code{QAG} results shown.}
\label{tab.rtc.checks1}\end{table}

\begin{table}\small\begin{tabular}{|>{$}c<{$}|>{$}p{.82\textwidth}<{$}|}
\hline
\langle\Psi,Q_{B}\Psi\rangle^{(2)}&0\\
\langle\Psi,\Psi*\Psi\rangle^{(2)}&0\\
\langle\Psi,Q_{B}\Psi\rangle^{(3)}&0\\
\langle\Psi,\Psi*\Psi\rangle^{(3)}&0\\
\langle\Psi,Q_{B}\Psi\rangle^{(4)}&-(8.7\pm1.2)\E{-10}+(0\pm4.2\E{-14})C^{L}\\
\langle\Psi,\Psi*\Psi\rangle^{(4)}&0\\
\langle\Psi,Q_{B}\Psi\rangle^{(5)}&0\\
\langle\Psi,\Psi*\Psi\rangle^{(5)}&0\\
\langle\Psi,Q_{B}\Psi\rangle^{(6)}&(0.02\pm0.30)+(0\pm3.5\E{-2})C^{L}+(0\pm0.11)(C^{L})^{2}+(0.5\pm9.6)\E{-13}(C^{L})^{3}+(0\pm6.7\E{-2})C_{1}-(0.1\pm2.3)\E{-13}C^{L}C_{1}+(0\pm1.1\E{-2})C_{0}+(0.1\pm1.5)\E{-14}C^{L}C_{0}\\
\langle\Psi,\Psi*\Psi\rangle^{(6)}&0\\
\hline
\end{tabular}
\caption[Deterministic evaluation of the action for the  
rolling tachyon]{Deterministic evaluation of the action for the  
rolling tachyon.  Kinetic and 
cubic terms are found separately as a consistency check.  
Superscripts represent the order in $\lambda$. 
\code{Cuhre}/\code{QAG} results shown.}\label{tab.rtc.checks2}
\end{table}


\clearpage


%

\bibliographystyle{JHEP}
\bibliography{refs}

\end{document}